\pdfoutput=1
\documentclass[twocolumn]{aastex631}
\usepackage[utf8]{inputenc}

\usepackage{graphicx}
\usepackage{mathtools}
\usepackage{multirow}
\usepackage{epstopdf}
\usepackage{amsmath}
\usepackage{hyperref}

\usepackage{tikz}
\usetikzlibrary{positioning}

\usepackage{float}

\newcommand{\SK}{$\widehat{SK}$}
\newcommand{\SKs}{$\widehat{SK}\,$}
\newcommand{\wfir}{$W_{FIR}$}

\received{2022 May 2}

\revised{2022 June 22}

\accepted{2022 July 3}

\begin{document}
\title{Simulating Spectral Kurtosis Mitigation Against Realistic RFI Signals}

\author[0000-0001-9968-5334]{E. Smith}
\affiliation{Department of Physics and Astronomy \\
West Virginia University \\
White Hall, Box 6315, Morgantown, WV 26506}
\affiliation{Center for Gravitational Waves and Cosmology \\
West Virginia University \\
Chestnut Ridge Research Building, Morgantown, WV 26505}

\author[0000-0001-5229-7430]{Ryan S.\ Lynch}
\affiliation{Green Bank Observatory (GBO) \\
155 Observatory Road, Green Bank, WV 24944}

\author[0000-0001-7996-7860]{D.J. Pisano}
\affiliation{Department of Physics and Astronomy \\
West Virginia University \\
White Hall, Box 6315, Morgantown, WV 26506}
\affiliation{Center for Gravitational Waves and Cosmology \\
West Virginia University \\
Chestnut Ridge Research Building, Morgantown, WV 26505}
\affiliation{Department of Astronomy\\  
University of Cape Town \\
Private Bag X3 \\
Rondebosch 7701, South Africa}


\begin{abstract}
    We investigate the effectiveness of the statistical radio frequency interference (RFI) mitigation technique spectral kurtosis (\SK) in the face of simulated realistic RFI signals. \SKs estimates the kurtosis of a collection of $M$ power values in a single channel and provides a detection metric that is able to discern between human-made RFI and incoherent astronomical signals of interest. We test the ability of \SKs to flag signals with various representative modulation types, data rates, duty cycles, and carrier frequencies. We flag with various accumulation lengths $M$ and implement multi-scale \SK, which combines information from adjacent time-frequency bins to mitigate weaknesses in single-scale \SK. We find that signals with significant sidelobe emission from high data rates are harder to flag, as well as signals with a 50\% effective duty cycle and weak signal-to-noise ratios. Multi-scale \SKs with at least one extra channel can detect both the center channel and side-band interference, flagging greater than 90\% as long as the bin channel width is wider in frequency than the RFI.
\end{abstract}

\section{Introduction}
\label{sec:intro}

In the face of ever-increasing human-made radio frequency interference (RFI) and larger datasets, there exists a need for new techniques of automated RFI mitigation. While policy groups such as the International Telecommunication Union (ITU) and other national organizations set standards for small protected radio astronomy bands scattered across the spectrum, astronomers are more often than not collecting data in unprotected bands that are rife with RFI.
Galactic and extragalactic imaging in the lower half of L-band (1000-1400 MHz) suffer from imaging artifacts produced by GPS and communication satellite transmissions \citep{hess2019chiles}, and high-redshift HI and Epoch of Reionization surveys have to deal with air traffic radar, satellite TV/phone services, and FM radio broadcasts \citep{offringa2015low,hunt2016search,fisher2005mitigation}. Non-astronomical science such as radio sensing for meteorologic and geophysical services compete with RFI produced by ground-based transmitters or reflections of space-based transmitters. \citep{andrews2021properties}.

As RFI encroaches more and more into radio bands of scientific interest, astronomical signals will increasingly be overlapped by harmful interference, so it is also important to devise and test RFI mitigation techniques that can detect and remove human-made signals without negatively impacting scientific data. Furthermore, modern spectrometers are handling incredibly large amounts of data per second, especially those serving as backends for radio telescope arrays or phased array feeds, and this data has to be averaged down or reduced in some way in order to make long-term storage feasible. Many modern RFI mitigation techniques happen in the post-averaging phase after the data leaves the spectrometer, which can have their own set of drawbacks \citep{offringa2013lofar}. For example, if a low duty cycle interferer affects a longer averaged scan or a very narrow signal saturates a much wider frequency channel, there may be a significant amount of clean data that is thrown out along with the RFI, as is documented for the Canadian Hydrogen Intensity Mapping Experiment \citep[CHIME;][]{mirhosseini2020high}. Thus, it is important to characterize online RFI mitigation techniques that can work on un-averaged data inside the spectrometer to increase their effectiveness \citep{baan2011rfi} while ensuring the extra computation involved is optimized enough to keep up with the data rate and not bottleneck the spectrometer.

Kurtosis as a statistical measure of a data set can be traced back to \citet{pearson1905fehlergesetz} and \citet{dwyer1983detection}, with a formalized definition in \citet{antoni2006spectral} and is extensively used for detecting failures in vibrating and rotating machines \citep{raad2008indicators,mcinerny2003basic,tian2015motor}, among other similar engineering problems. Significant work has been done to improve this methods ability to detect transient events in diagnostic data, including envelope kurtosis\citep{barszcz2011novel}, kurtosis ratio\citep{vass2008avoidance}, and kurtosis wavelet analysis\citep{wang2013energy}. This method has a wide range of applications, including detection of both transient and persistent RFI in radio astronomy.

In this paper we explore the feasibility of the spectral kurtosis estimator (\SKs) \citep{nita2016spectral}, which determines the Gaussianity of channelized voltages by assessing the distribution of their associated squared power values per spectral channel over time. It is used to flag time ranges of channels that fall outside of a certain acceptable threshold, and it is a computationally simple method that has been proven to mitigate RFI in real-time inside the spectrometer \citep{gary2010wideband,nita2016eovsa}.

Real, incoherent, and stationary astronomical signals are expected to have Gaussian noise statistics with a certain mean and standard deviation, while having no skew and no excess kurtosis. Kurtosis, also known as the fourth moment of a dataset, is the sharpness or flatness of a Gaussian distribution. When coherent RFI signals are present in the data, they can cause kurtosis that differs from what is expected for Gaussian noise, allowing astronomers to set a threshold beyond which the data is flagged. The \SKs estimator, on the other hand, is designed to approximate the kurtosis of the channelized voltages using the squared power values, which is a simpler calculation. The main advantage of this method is its ability to detect and flag RFI signals while preserving astronomical signals. Since it is developed specifically for online usage, \SKs allows for flagging channels at high time resolution and raising the effective integration time for a given observation wall-clock time, as well as lowering the achievable system noise.


\SKs can be categorized as a higher order statistical RFI detection method. However, the statistical qualities of different RFI signals can range fairly wide and may react differently to \SKs detection. A typical radio observatory can have a large collection of incoming human-made radio signals corrupting data. These signals serve different purposes and have different properties. For example, the Iridium global satellite phone constellation\footnote{\url{https://www.sigidwiki.com/wiki/Iridium}} is an ever present phase-shift keyed signal at a maximum data rate of 50 kbps across 12 frequency channels from 1620 - 1626 MHz, and the terrestrial Bedford, NC aviation radar sends two half-second pulses every roughly 12 seconds from a fixed spot on the horizon at the Green Bank Observatory \citep{fisher2001a}. While these sources are known and well documented, there can also be a plethora of undocumented or unknown signals that are not necessarily detected with the same effectiveness.

In this paper, we explore the flagging efficacy of \SKs on simulated RFI with various statistical properties to determine how well each is detected. We aim to determine the optimal setup for an unknown RFI environment that will detect the most RFI possible, as well as provide insight into how to detect a specific RFI source with known qualities.

In Section \ref{sec:methods}, we describe how flagging is applied and evaluated, then describe the process of creating simulated datasets in Section \ref{sec:data}. Results are shown via flagging charts in Section \ref{sec:results}, and these are contextualized in Section \ref{sec:disc}.

\section{Methods}
\label{sec:methods}

The SK estimator of a set of complex-valued data can be calculated with the formula

\begin{equation}
    \widehat{SK} = \frac{MNd + 1}{M - 1}\Big(\frac{M \cdot S_2}{S_1^2}-1\Big)
    \label{func:SK_def}
\end{equation}

with $M$ as the accumulation length of power spectral density (PSD) values, $d$ as an empirically determined shape factor, and $N$ as the number of subsequent spectra that are averaged together inside the spectrometer before being dumped to a data product. \citep{gary2010wideband,nita2016spectral,nita2010generalized,nita2010statistics,nita2007radio,nita2016eovsa} In this work, we use $N = d = 1$, as we are not averaging any spectra before \SKs mitigation, and a zero-mean Gaussian noise profile corresponds to $d=1$. $S_1$ and $S_2$ are defined per spectral channel $k$ as

\begin{equation}
\begin{split}
    S_1(k)=\sum_{i=1}^M\Big(\sum_{j=1}^NP_j\Big)_i \\
    S_2(k)=\sum_{i=1}^M\Big(\sum_{j=1}^NP_j\Big)_i^2
\end{split}
\label{func:S1_S2_def}
\end{equation}

where $P_j$ are the power spectral density values. \SKs is a well-defined estimator for the Gaussianity of the raw time-series voltages. For RFI-free active white Gaussian noise voltages, The statistical distribution of \SKs is a Pearson type III curve, which is centered at 1, has a left skew, and has a spread determined almost entirely by the accumulation length $M$. We find the upper threshold for acceptable \SKs values by solving for the intersection of the cumulative distribution function (CDF) and 0.0013499, which is the probability of false alarm corresponding to a 3$\sigma$ detection level. The lower threshold is computed by finding the intersection of 0.0013499 and the function 1 - CDF. These are more precise than using just the \SKs standard deviation and lead to slightly asymmetric thresholds around 1. Spectral Kurtosis is run individually on each channel on $M$ spectra at a time. $M$ should be large enough to avoid errors from small statistics ($>$ $\sim$200-300), but this comes with the caveat that larger accumulations leads to coarser time resolution. An appropriate value of $M$ is contextual, depending on the data rate of the spectrometer, time resolution of single spectra, and intermittence of expected RFI.


The \SKs estimator formula described above in Equation \ref{func:SK_def} can be derived from the spectral variability $V_k$:

\begin{equation}
V_k = \frac{\sigma_k^2}{\mu_k^2}
\label{eq:var_mean2}
\end{equation}

following the steps in \citep{nita2007radio}, with $\mu_k$ as the mean and $\sigma_k^2$ as the variance of power values in frequency channel $k$. It is weak to signals with $\sim50\%$ duty cycle because the squared mean and variance of the power values are similar and $V_k$ approaches unity \citep{nita2010generalized}. In this case, the standard \SKs implementation does not detect RFI. However, There is a variant of \SKs called multi-scale \SKs that averages rolling windows of adjacent time-frequency pixels in the arrays of $S_1$ and $S_2$ and allows us to bypass this weakness to $50\%$ duty cycle signals \citep{gary2010wideband}. For a multi-scale \SKs bin width of $m$ channels and $n$ \SKs time bins, our new $S_1$ and $S_2$ are

\begin{equation}
\label{eq:MS_S1_S2}
\begin{split}
    S_{1,ik}^{mn}=\frac{1}{m\cdot n}\sum_{j=0}^{m-1}\sum_{l=0}^{n-1}S_{1,k+j,i+l} \\
    S_{2,ik}^{mn}=\frac{1}{m\cdot n}\sum_{j=0}^{m-1}\sum_{l=0}^{n-1}S_{2,k+j,i+l}
\end{split}
\end{equation}

for channel index $k$ and time index $i$. The combined duty cycle $d$ of signals in each multi-scale bin is now made up of the duty cycles of signals in the single-scale pixels,

\begin{equation}
\label{eq:ms_dc}
d_{ik}^{mn} = \frac{1}{m\cdot n} \sum_{j=0}^{m-1}\sum_{l=0}^{n-1}d_{k+j,i+l}
\end{equation}

so if adjacent pixels have a duty cycle closer to 0\% (no RFI) or 100\% (continuous RFI), the resulting effective duty cycle can be pulled away from 50\%. \SKs is applied with the new $S_{1,ik}$ and $S_{2,ik}$, and if the multi-scale bin is found to have RFI, all of the original single-scale time-frequency pixels within the bin are flagged.

The simulation process starts by generating the time samples of an RFI signal summing with additive white gaussian noise. The RFI is ramped up from zero amplitude over the course of the scan to help discern how bright the signal has to be before \SKs detection. Specific details about the signal generation process are outlined in Section \ref{sec:data}. The data stream is then channelized via a poly-phase filterbank routine \citep{price2021spectrometers}, which uses 24 taps and a Hanning window to smooth the filter shape, as this mimicks the VEGAS spectrometer at the Robert C.Byrd  Green Bank Telescope \citep[GBT;][]{prestage2015versatile}. The resulting data is reshaped and squared, resulting in a 2-dimensional array of un-averaged pseudo-power spectra.

Next, we apply both single-scale and multi-scale \SKs mitigation to the data. First, we split up the data into chunks of $M$ spectra and compute \SKs using Equation \ref{func:SK_def}. The outputs are 2-dimensional arrays of the \SKs values, averaged power, and flagging mask - and the time axis of these arrays is shortened from the native data set size by a factor of $M$. The flagging mask is a boolean array. A similar function then performs multi-scale \SKs given the bin size and outputs the corresponding arrays. As described above, we apply a rolling average window to the $S_1$ and $S_2$ accumulated power and squared power values, then evaluate \SKs using the same Equation \ref{func:SK_def}. Since multi-scale \SKs uses this larger rolling window, we flag every single-scale pixel contained in each multi-scale bin.

In order to compare flagging efficacy against different mitigation schemes and RFI characteristics, we need a standard way to generate flagging percentages. Since the signals can have an appreciable sidelobe structure, a comparison mask is used to determine where the RFI is strong. We channelize the RFI independently from the noise and convert into decibels using the noise as reference.
Whenever the signal is more than -10dB compared to the noise level, we consider RFI to be present. -10dB is chosen as a medium value between overflagging and missing RFI. We call our \SKs mask $M_{SK}$ and the comparison mask $M_{dB}$. The number of true positive points is defined as the number of pixels where both the \SKs and comparison masks are true, or the intersection $M_{SK} \cap M_{dB}$. The number of true negative points, or where the \SKs flag correctly identified no RFI, is the intersection of when both masks are untrue, denoted by a prime: $M_{SK}' \cap M_{dB}'$. False positives are when our \SKs mask incorrectly flags noise, and false negatives are when we miss flagging RFI. We define true positive rates (TPR) and false positive rates (FPR) following these definitions:  


\begin{equation}
\begin{split}
    \mathrm{N_{TP}} = \mathrm{num}(M_{SK} \cap M_{dB} ) \\
    \mathrm{N_{TN}} = \mathrm{num}(M_{SK}' \cap M_{dB}' ) \\
    \mathrm{N_{FP}} = \mathrm{num}(M_{SK} \cap M_{dB}' ) \\
    \mathrm{N_{FN}} = \mathrm{num}(M_{SK}' \cap M_{dB} ) \\
    \mathrm{True\, Positive\, Rate} = 100 \cdot \frac{N_{TP}}{N_{TP}+N_{FN}}\\
    \mathrm{False\, Positive\, Rate} = 100 \cdot \frac{N_{FP}}{N_{FP}+N_{TN}}
\end{split}
\end{equation}

This procedure is done for both the single-scale \SKs flagging mask and the union of the single-scale and multi-scale \SKs mask, in order to determine if the multi-scale \SKs has any significant advantages. These percentages are then recorded for each simulation run. 

\begin{figure}[ht]
\centering
\includegraphics[width=0.7\linewidth]{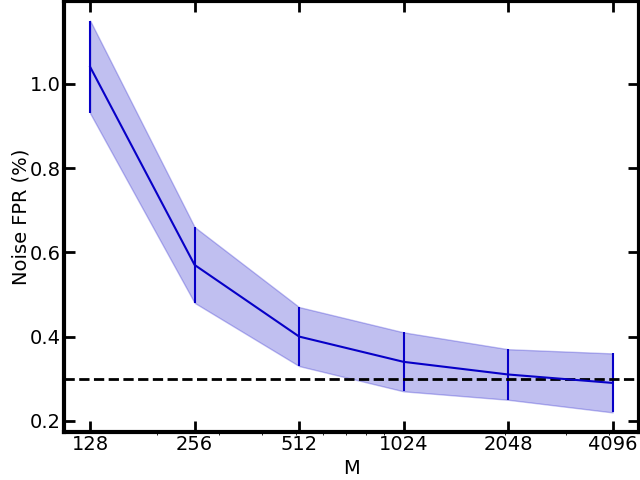}
\caption{False Positive Rates of \SKs on signal-free noise, depending on $M$.}
\label{fig:fpr_noise}
\end{figure}

To contextualize the false positive rates, we run the simulations with no RFI present. A dependence on $M$ was found, shown by the results in Figure \ref{fig:fpr_noise}. Any FPR values higher than the ones shown indicate that more than the acceptable amount of clean data was flagged. Multi-scale \SK was not found to flag any clean data. For large $M$, we approach the Gaussian 3$\sigma$ level of 0.3\%, which is marked on the graph with a horizontal dashed line.

For each simulation run, a spectrogram of the original unflagged data is saved, as well as spectrograms of the single-scale and single-scale plus multi-scale \SKs flags applied to the data. In addition, scatter plots are generated showing log$S_1$ vs. \SKs for each time-frequency pixel. These plots reveal differences in the way RFI is detected based on various characteristics, including the strength of the signal, the modulation type, and its presence in the center frequency or sideband channels. At the end of the simulation run, only the plots and flagging percentages are saved.

As mentioned above, the simulation process is done for a wide range of different RFI characteristics. For each signal characteristic, a realistic or interesting range of values is examined and compared in a series of 2-dimensional experiments. Table \ref{tbl:RFI_chars} shows the various possible values for the different parameters.
 
The data rates are chosen to balance realistic values while also being at relatable speeds compared to the data rate of our spectrometer. At the low end of 1 kilo-symbol per second (ksps), the symbol changes only after $\sim10^3$ spectra have been collected, meaning that the signal acts more like an un-modulated continuous carrier signal over long periods of time. Lowering the data rate even further will not change the flagging efficacy. At the high end of 100-200 ksps, the modulation can happen on time-scales shorter than the time it takes to collect a single spectrum's worth of time samples. It is interesting to see what happens to the \SKs flagging efficacy when this sampling breaks down, but as 200 ksps is a data rate typically seen only in wired connections, it is not useful to push this upper bound further. Keep in mind that the corresponding bit rate depends on how many bits per symbol there are, so a 4-level modulation scheme running at 200 ksps corresponds to 800 kilo-bits per second.
 
Values for $M$ run from 128 to 4096 in powers of 2, to balance the time resolution of \SKs flagging against having a statistically significant amount of spectra. 128 is picked as the low end, since it may be difficult to determine the kurtosis of less than 128 values but also yields the fastest time resolution. We also expect the errors to go as $1/\sqrt{M}$, which becomes greater than 10\% as M drops below 100. $M$ = 4096 provides the most stable statistical measures, but the slowest time resolution. Higher $M$ was difficult to accomplish given time and computing resource restraints.
 
Multi-scale \SKs bin shapes are denoted by MS-$mn$ with $m,n$ from equation \ref{eq:MS_S1_S2} where the first dimension is the size of each bin in channels and the second is the size of each bin in single-scale \SKs bins, that is, chunks of $M$ power values.  MS-24, for example, denotes multi-scale \SKs with a bin size of 2 channels and 4 single-scale \SKs time bins. Various shapes are chosen with sides of 1,2, and 4.
  
The central frequency of the signal is chosen to either center on a PFB channel or be slightly off to the side, to investigate how the flagging efficacy is affected by any differences in how \SKs treats the center channel and sidelobes.

We also explore the difference in the cutoff values of the FIR windows, as it has interesting effects on \SKs flagging. Wireless transmitters will apply a smoothing window to lower the signal bandwidth, both for energy-efficiency reasons and to comply with telecommunication laws. However, it is not easy to narrow down a most accepted way to do this, as every telecommunications system does something different and their methods are not easily searchable, so we take two different cutoff values of $1/W_{FIR}$ and $4/W_{FIR}$, where the latter value results in wider signals. The total FIR window size is set to be 20\% that of each symbol, to balance the smoothing amount against a receiver's ability to recover the data. Too much smoothing will reduce the sidelobes even further, but could make it impossible to detect the original data sequence. The intended receiver for the signal has to be able to discern the bits of the transmitted signal, and if the transition between bits is too smooth, it may become impossible to recover the original bit stream. 
 

\begin{deluxetable*}{cccccc}[ht]
\tabletypesize{\footnotesize}
\tablecolumns{6}
\tablewidth{60pt}
\tablecaption{Values for the different RFI and flagging parameters.\label{tbl:RFI_chars}}
\tablehead{
\colhead{Data Rate} & \colhead{Duty Cycle} & \colhead{\SKs $M$} & \colhead{MS shape} & \colhead{Channel Center} & \colhead{FIR cutoff} \\ 
\colhead{(ksps)} & \colhead{(\%)} & \colhead{(\# spectra)} & \colhead{} & \colhead{} & \colhead{}} 

\startdata
1   & 5     & 128   & Single-scale & 120    & $1/W_{FIR}$ \\
4   & 10    & 256   & MS-12       & 120.25  & $4/W_{FIR}$ \\
20  & 15    & 512   & MS-21       & 120.5  &              \\
100 & ...   & 1024  & MS-22       &        &              \\
200 &       & 2048  & MS-24       &        &              \\
    & ...   & 4096  & MS-42       &        &              \\
    & 95    &       &             &        &              \\
    & 100   &       &             &        &              \\
\enddata
\end{deluxetable*}

To summarize the tests of different RFI characteristics and mitigation schemes on flagging efficacy, we compare for each modulation type:
\begin{enumerate}
    \item Signal data rate to \SKs $M$, for single-scale and multi-scale \SK
    \item Signal data rate to duty cycle, for single-scale and multi-scale \SKs
    \item Signal data rate to multi-scale \SKs bin shape
    \item Signal data rate to multi-scale \SKs bin shape
    \item Duty cycle to a range of multi-scale \SKs shapes
    \item Each of these with carrier frequencies offset by a quarter or half channel
\end{enumerate}


In order to generate flagging plots as shown in Section \ref{sec:results}, a script was written to simulate every combination of two parameters in Table \ref{tbl:RFI_chars} one hundred times each to generate a statistically significant population of flagging results from which means and variances can be calculated.


\section{Data}
\label{sec:data}

We developed a collection of signal generating functions to create 1-dimensional time sample streams of RFI with a varying range of characteristics, including modulation type, amplitude, duty cycle, duty cycle period, and data rate. The data were generated to have the same or similar bandwidth to test observations of pulsars taken with the GBT, which had a bandwidth of 800MHz with 4096 channels. With a sample rate of 50 MHz and 256 channels, we match this time/frequency resolution, but with far smaller data sets to analyze and store in memory. The only requirement is that there are enough channels to have an RFI-contaminated and an RFI-free part of the data set. In the spectrograms shown below, only the relevant 40 channels surrounding the RFI are shown. With this time/frequency resolution, each spectrum has a time resolution of 5.12$\mu$s. Picking $M=512$ with 300 \SKs blocks will generate a dataset with 153,600 spectra, which corresponds to 0.786 seconds worth of data. Although this is much shorter than a typical observation length, it is enough to recreate a representative RFI signal and analyze it.

Using online resources such as the SatNOGS database\citep{nicolas2021satnogs}, the Signal Identification Wiki\footnote{\url{https://www.sigidwiki.com/wiki/Signal_Identification_Guide}}, and the World Meteorological Organization's OSCAR\citep{world2021observing} tool to explore and document possible RFI signals, we found that many current satellite communications use some form of phase-shift keying (PSK), amplitude-shift keying (ASK), or frequency-shift keying (FSK). These modulation types encode a specific information stream of bits along an electro-magnetic wave at a specific carrier frequency or frequencies. For our simulations, they are used in conjunction with realistic or interesting data rates, duty cycles, and duty cycle periods given the context of the sampling rate and channel number of the simulated data set. For example, a quadrature phase-shift keyed signal is created by complex-valued wave at four different phases ($0\degr,90\degr,180\degr$, and $270\degr$), one for each bit. The receiver then knows to interpret each phase as a 2-bit word and process the data accordingly.

The general form of each RFI generation function is very similar.\footnote{Functions were modeled after examples found at \url{http://cyclostationary.blog}.} The inputs include the number of bits, the symbol rate (in kilo-symbols per second), the carrier frequency (in MHz), and the sampling rate (in MHz). Given an \SKs $M$ value, the number of channels, and the number of different \SKs bins, the program derives how many bits to generate by dividing the total amount of time samples by the number of time samples per bit. Depending on the modulation type, more inputs may be required - such as an extra carrier frequency for FSK signals or a bias factor to determine the difference in power levels between each bit for ASK signals. Given the sampling rate and the symbol rate, the function defines how many time samples are in each bit and generates a random bit sequence according to the modulation type and size of each word. For example, a binary modulation scheme would generate 0's and 1's for bits while a quadrature scheme generates 0's, 1's, 2's, and 3's. Next, a complex carrier frequency signal is generated and modified using the modulation type and the bit sequence. For the binary phase-shift keyed (BPSK) signals, the bits are transformed from a sequence of 0's and 1's to a sequence of $\pm$1's and multiplied with the carrier signal, flipping its sign for either phase. ASK signals with $N$ bits per symbol and bit sequence $B$ have the amplitude $v_i$ of the carrier wave modulated by

\begin{equation}
v_i = \frac{2B}{2^N-1}-1
\end{equation}

So, a 2 bit ASK signal has amplitudes of -1, -0.33, +0.33, and +1. For FSK modulation, the signal has to be generated such that the phase stays constant across frequency shifts. A function that acts as a voltage-controlled oscillator aids in creating these frequency-shift keyed signals. In addition, the bit sequence of each signal is run through a rectangular FIR filter before modulating the carrier frequency. This is done by generating a sinc function with a width of 20\% that of each symbol, and a cutoff of either 1 or 4 times the reciprocal of the width for a wider or narrower signal ($1/W_{FIR}$ or $4/W_{FIR}$). That sinc function is then convolved with the bit sequence. Each RFI generator outputs a 1-dimensional numpy array of complex voltage samples, as well as the bit sequence used.

Many RFI transmissions have some form of duty cycle, or period over which time the signal spends some fraction on and off. This is usually done to conserve energy or bandwidth, or allow the receiver to decipher other frequency channels easier. Since it is known that \SKs responds to this signal characteristic in different ways \citep{nita2007radio}, we test the full range of duty cycles from 5\% to 100\% by setting a period of 1 millisecond and setting some fraction of the data to 0 for however many duty cycle periods there are. This period of 1 millisecond is chosen as a middle ground for typical rate that digitally encoded signals may turn on and off for. It is much longer than the spectral rate but on similar timescales to the \SKs bin rate. For example, when we use $M=512$, there are 2.62144 duty cycles contained in each independent \SKs calculation. This odd ratio is by design in our simulations, as the duty cycle of the transmitter does not know or care whether our radio astronomy spectrometer works at a fortuitously similar rate (e.g. 2/3 or 4/5). However, this means that the intrinsic duty cycle we give our signal may manifest as a different effective duty cycle after passing through the PFB and \SKs mitigation. In our example setup then, a 20\% intrinsic duty cycle signal could have anywhere between 2 and 3 full sets of 20\% ON cycles, for an effective duty cycle ranging from 15.3\% to 22.9\%. While this may make for somewhat inaccurate duty cycles in our simulation results, we expect the duty cycles of realistic signals to behave in a similar way with \SKs in an online spectrometer. Using a larger $M$ or a smaller duty cycle period will mitigate this and allow the effective duty cycle to match the intrinsic more easily.

\SKs should not flag incoherent astronomical signals, and we test this by mimicking a spectral line with a Gaussian profile. On the order of $10^4$ small continuous wave signals were summed together, with frequencies determined by a Gaussian distribution centered on a PFB channel and uniform random phase delays, in order to obfuscate any semblance of non-stationarity or coherence in the output signal.


\section{Results}
\label{sec:results}


The following subsections are ordered by experiment, with each describing the results from all the tested modulation types. When comparing flagging rates, we round to the nearest whole percent and classify differences in absolute flagging percentages as follows: $<$5\% is an insignificant change, 5-10\% is a slight change, 10-20\% is a moderate change, and $>$20\% is a major change. For example, in Figure \ref{fig:f_ksps_SKm}b), there is only a slight difference in flagging rate ($<$10\%) between $M=128$ and $M=4096$ \SKs at a symbol rate of 4 ksps, but there is a major difference ($\sim$40\%) for a 20 ksps signal.

\subsection{Signal Data Rate to \SKs $M$}
\label{sec:ksps-m}

\begin{figure*}[ht!]
    \centering
    \begin{tabular}{ c c }
    \includegraphics[width=0.32\linewidth]{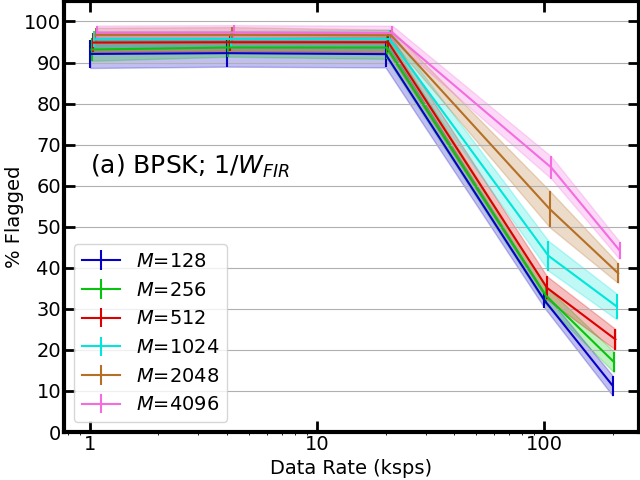}
    &
    \includegraphics[width=0.32\linewidth]{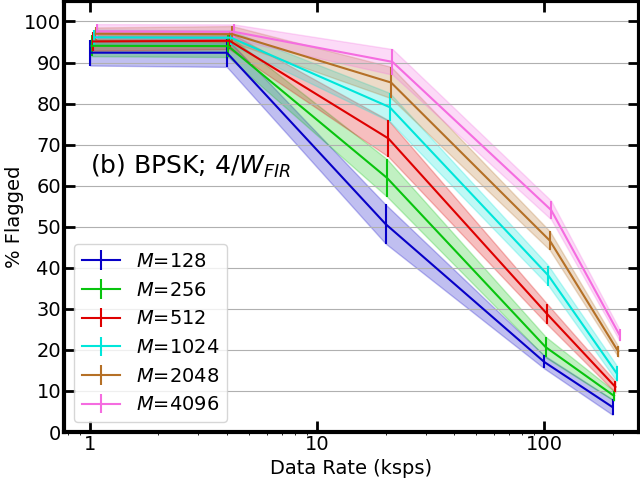}
    \\
    \includegraphics[width=0.32\linewidth]{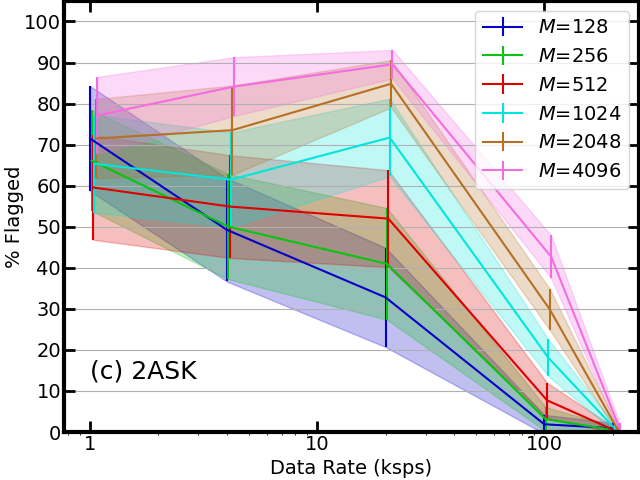}
    &
    \includegraphics[width=0.32\linewidth]{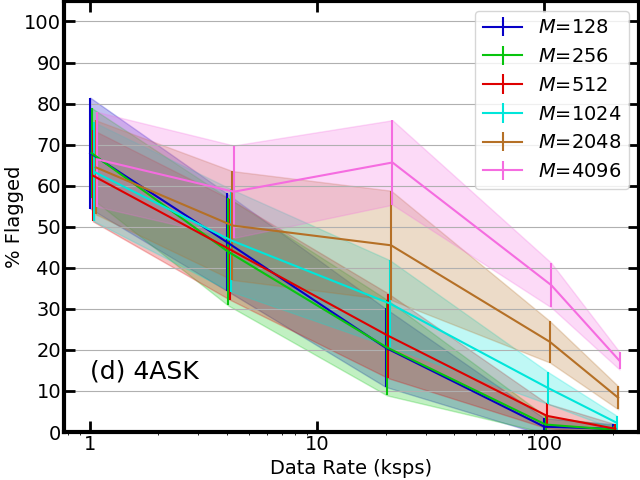}
    \\
    \end{tabular}
    \caption{Flagging percentages for various values of $M$ compared to the data rate of the RFI signal. a) BPSK modulation with a $1/W_{FIR}$ cutoff. b) BPSK modulation with a $4/W_{FIR}$ cutoff. c) 2-bit ASK modulation with a $1/W_{FIR}$ cutoff. d) 4-bit ASK modulation with a $1/W_{FIR}$ cutoff.}
    \label{fig:f_ksps_SKm}
\end{figure*}

\begin{figure*}[ht!]
    \centering
    
    \begin{tabular}{ccc}
    \includegraphics[width=0.32\linewidth]{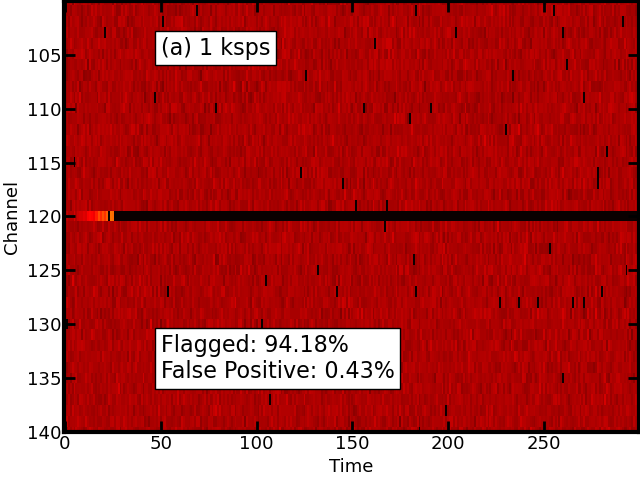}
    &
    \includegraphics[width=0.32\linewidth]{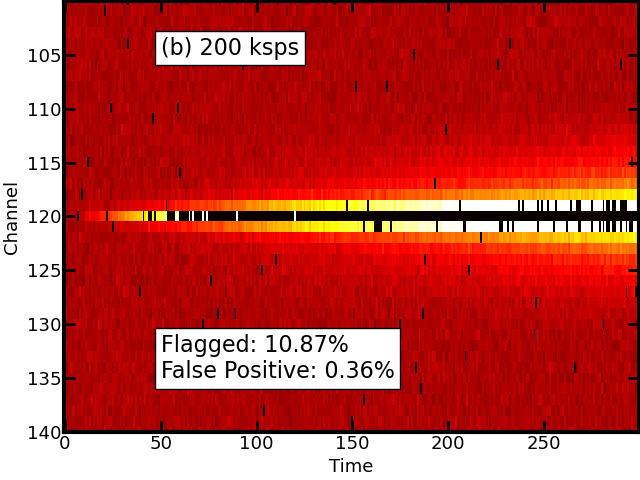}
    &
    \multirow{2}{*}[2in]{
    \includegraphics[width=0.06\linewidth]{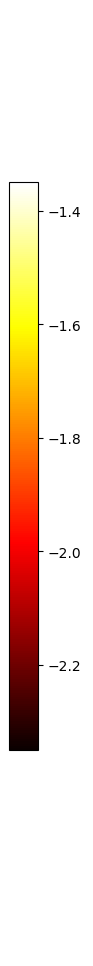}
    } \\
    \includegraphics[width=0.32\linewidth]{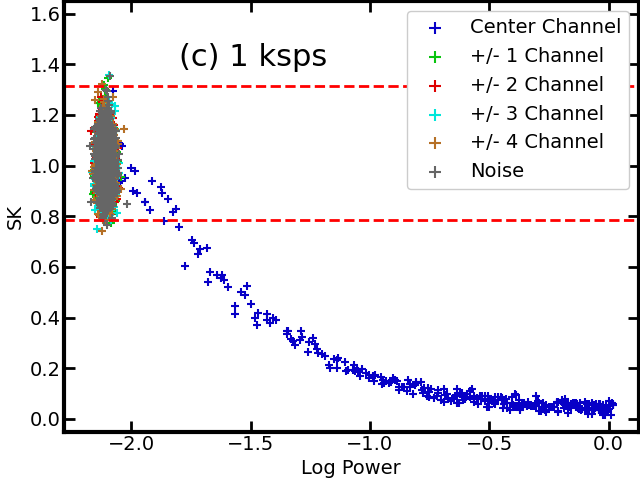}
    &
    \includegraphics[width=0.32\linewidth]{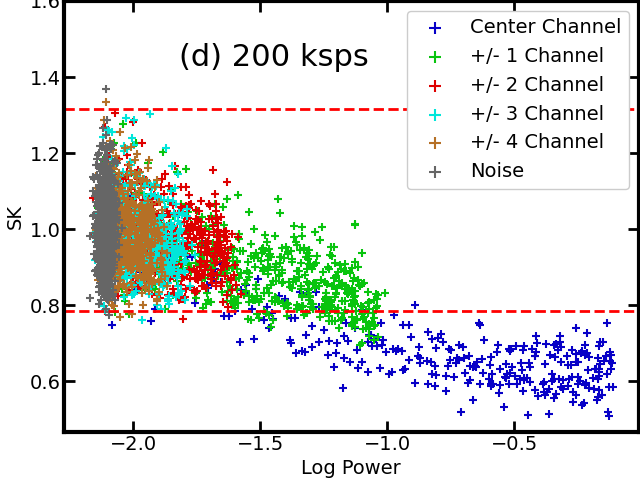}
    &                  
    \end{tabular}
    \caption{Flagged RFI and \SKs vs. log power for a 1 ksps and 200 ksps BPSK signal. As the data rate rises, the sidelobes of the signal become much more powerful but fail to be detected by \SK. In panels c and d, the red lines denote the acceptable \SKs thresholds. Any points that don't lie between these thresholds are flagged as RFI.}
    \label{fig:ksps_SKM_1}
\end{figure*}

\begin{figure}[ht!]
    \centering
    \includegraphics[width=0.7\linewidth]{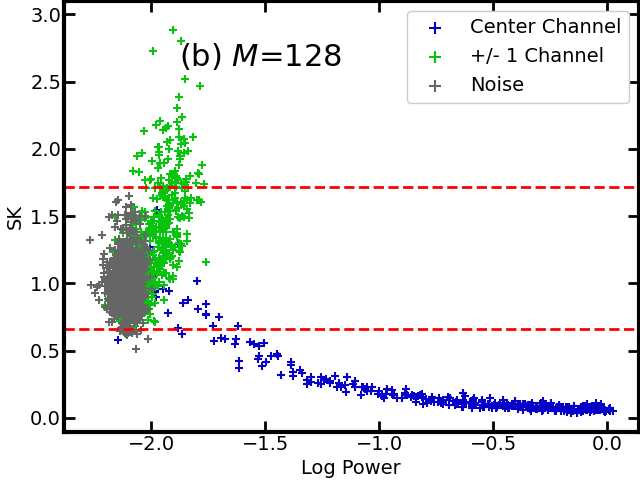}
    \includegraphics[width=0.7\linewidth]{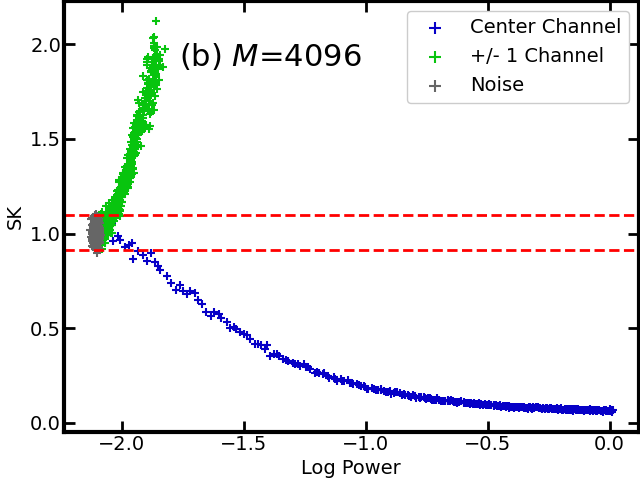}
    \caption{\SKs vs.\ log power for (a) $M=128$ and (b) $M=4096$. As in the above figures, the center channel is colored blue, the first sidelobe channel in green, and noise in gray. As $M$ rises, the \SKs values of the weak sidelobes move away from the gray noise more dramatically, leading to cleaner flagging. This data was generated at 20 ksps, which has noticeable sidelobes extending 1 channel on either side of the center channel. \SKs thresholds are outlined in red.}
    \label{fig:ksps_SKM_2}
\end{figure}

Figure \ref{fig:f_ksps_SKm} shows the effectiveness of \SKs mitigation for different $M$ values across a range of data rates. The two results we draw are that higher data rates lead to less flagging, but a higher $M$ does lead to higher TPR. To explain this, consider that the symbol sequence that modulates the carrier wave acts like a square wave. For higher data rate signals, this square wave has a smaller and smaller period. After channelization, this causes wider sidelobes which do not get detected as cleanly as the center channel. The faster the data rate, the larger the sidelobes, and the worse the flagging rate is for single-scale \SK. This is shown by the scatter plots of Figure \ref{fig:ksps_SKM_1}. For the 200ksps signal on the right side, the sidelobe emission is significantly higher in power than the noise, but do not depart from unity enough to be flagged as RFI. Larger values for $M$ subvert this weakness, as demonstrated by Figure \ref{fig:ksps_SKM_2}. For Figure \ref{fig:ksps_SKM_2}b, the \SKs values of data points in channels 119 and 121 (green) have a much more dramatic departure from the noise (gray), leading to easier flagging of dim sidelobes.

The other noticeable feature of Figure \ref{fig:f_ksps_SKm}a and b, for BPSK signals, is the flat flagging curve at lower data rates. It is more prominent on the top plot, which shows signals with a stricter $1/W_{FIR}$ cutoff frequency in the smoothing window. This happens because the signal data rates are low enough not to induce any sidelobes at our frequency resolution; in other words, the RFI is only one channel wide at data rates $<$20 ksps for a $1/W_{FIR}$ cutoff and $<$4 ksps for a $4/W_{FIR}$ cutoff. For these one-channel signals (which have a duty cycle of 100\%), \SKs flags almost all of the signal except for the faint portions, as seen in Figure \ref{fig:ksps_SKM_1}a. As this cutoff frequency only affects the sidelobe width and not the flagging effectiveness directly, we will continue to use the more strict $1/W_{FIR}$ cutoff frequency for the remainder of the results. QPSK signals are not shown in Figure \ref{fig:f_ksps_SKm}, as they are flagged along the same trends, although at slightly lower rates to BPSK signals, due to slightly larger sidelobes. 

ASK signals are also flagged better at lower data rates and higher $M$, but less overall and completely undetected at 200ksps, as shown in Figure \ref{fig:f_ksps_SKm}c) and d). 4-bit ASK signals are flagged slightly less, but only at higher $M$ - at the lowest $M$ value of 128, they are flagged very similarly. As $M$ rises, the disparity between the two bit rates rises. There is also a local maximum in data rates because the sidelobes start to get flagged more reliably before the center channel gets less flagged. Once the data rate gets too high, however, the signal completely escapes detection in all cases except for high $M$ with 4-bit ASK signals. This is likely due to how the effective duty cycle of an undersampled ASK signal appears to the \SKs algorithm, and this is explored more in Sections \ref{res:ksps-dc} and \ref{sec:disc-dc}.

\begin{figure*}[ht!]
    \centering
    \begin{tabular}{ c c c }
    \includegraphics[width=0.32\linewidth]{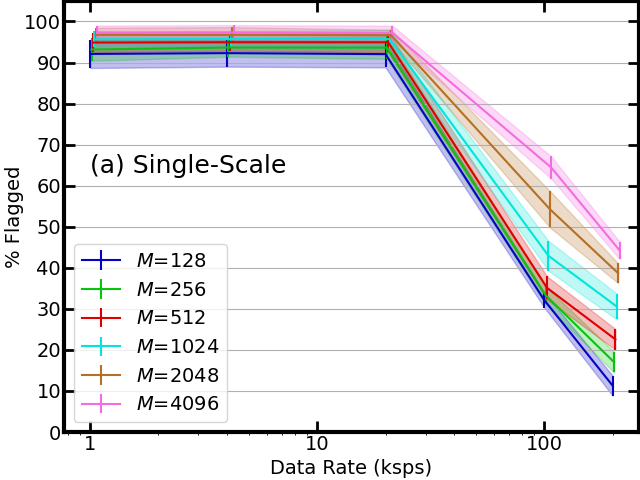} &
    \includegraphics[width=0.32\linewidth]{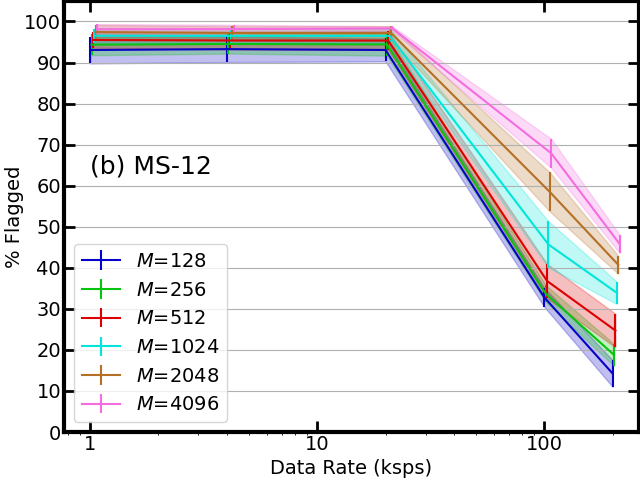} \\
    \includegraphics[width=0.32\linewidth]{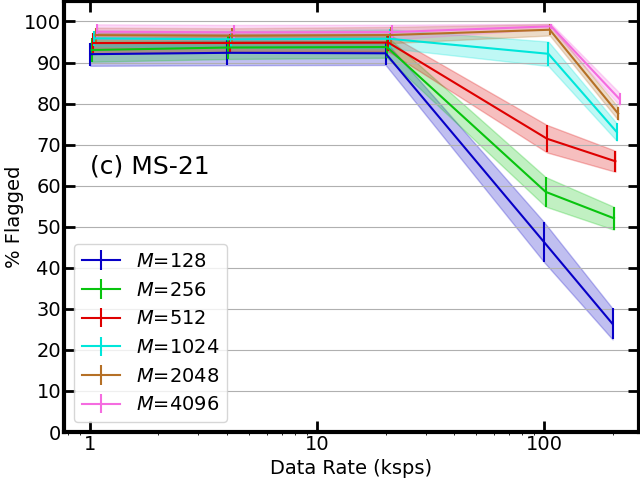} &
    \includegraphics[width=0.32\linewidth]{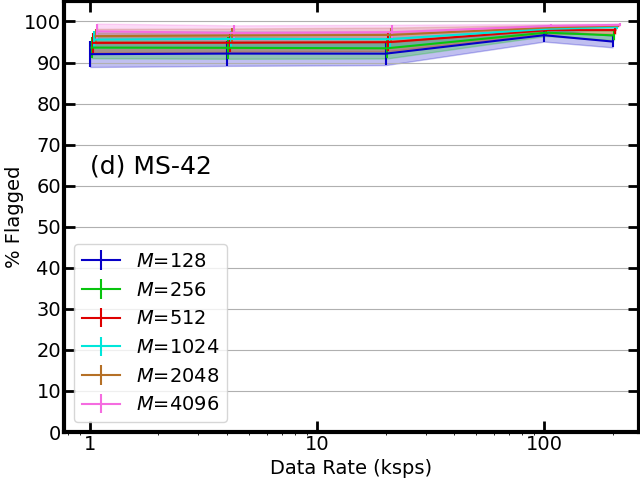} \\
    \end{tabular}
    \caption{Flagging percentages comparing data rate to $M$ for several multiscale \SKs bin shapes. a) Single-scale \SKs, b) MS-12, c) MS-21, d) MS-42.\newline}
    \label{fig:ksps_SKM_MSSK}
\end{figure*}

\begin{figure*}[ht!]
    \centering
    \begin{tabular}{ c c c }
    \includegraphics[width=0.32\linewidth]{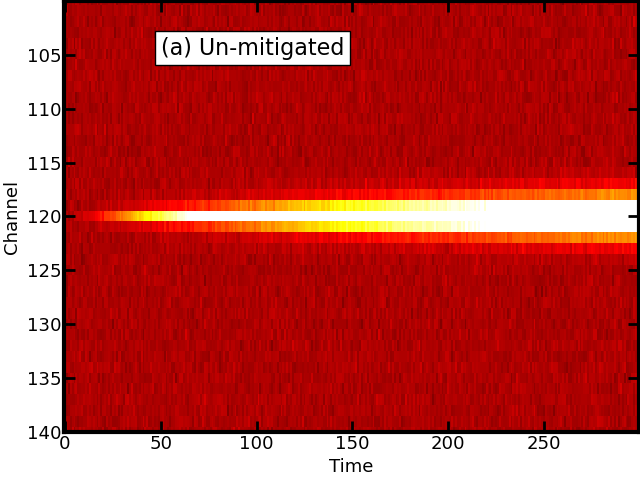} &
    \includegraphics[width=0.32\linewidth]{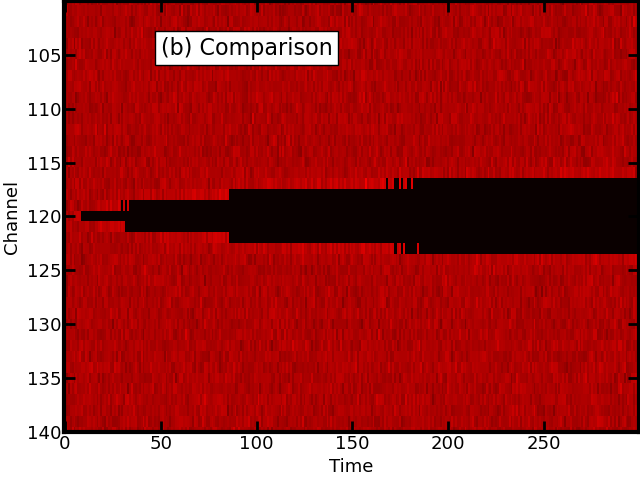} &
    \multirow{2}{*}[2in]{
    \includegraphics[width=0.06\linewidth]{misc/cbar.png} } \\
    \includegraphics[width=0.32\linewidth]{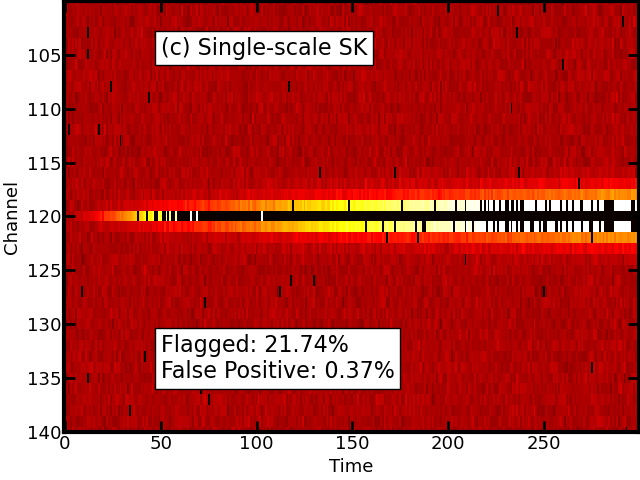} &
    \includegraphics[width=0.32\linewidth]{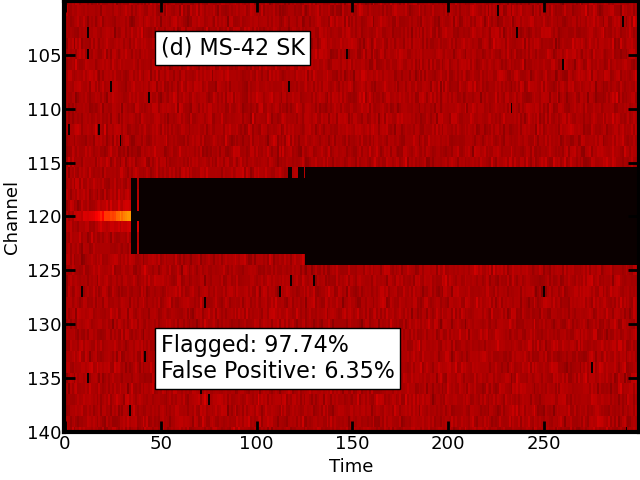} & 
    \end{tabular}
    \caption{Examples of flagging a 200 ksps BPSK signal with a $4/W_{FIR}$ cutoff frequency. (a) The unmitigated signal. (b) Comparison mask. (c) Single-scale \SK. (d) MS-42 \SK.}
    \label{fig:ksps_SKM_MS42}
\end{figure*}

We also chose a few select multi-scale \SKs shapes to see if some of the weaknesses in single-scale \SKs could be avoided. For BPSK signals, Figure \ref{fig:ksps_SKM_MSSK} shows the effect of MS-12, MS-21, and MS-42 multi-scale \SK on different $M$ and data rates. Comparing panels a and b, we can see that MS-12 does nothing to increase flagging; this will be expanded upon in following sections. Comparing panels c and d to a, however, we see that at higher data rates MS-21 \SKs performs somewhat better than single-scale \SK, especially at higher $M$. Including an extra channel in the multi-scale \SKs window allows the algorithm to detect some of the sidelobe emission. MS-42 dramatically raises flagging above the 90\% level for all data rates and values of $M$.
Figure \ref{fig:ksps_SKM_MS42} shows how the sidebands of a 200 ksps signal can be missed by single-scale \SKs but not by MS-42 \SK. In panel d, the multi-scale channel width is wide enough to capture all of the fainter sidelobe emission flagged by the comparison mask. Using a less strict 4/$W_{FIR}$ cutoff frequency instead of 1/$W_{FIR}$ allows the signal to become wider than the multi-scale \SKs width, and some of that sidelobe emission manages to escape detection.

The drawback of MS-42 \SK is that it can heavily overflag clean data. In the low data rate case, the false positive rates rise up to 2.6 - 2.9\%, almost 2 percentage points higher than the baseline noise FPR's of Figure \ref{fig:fpr_noise}. In these cases, MS-\SKs flags an extra three clean channels on either side of the center channel, even when the RFI doesn't actually have any sidelobes. This is apparent in comparing the left sides of Figures \ref{fig:ksps_SKM_MS42}b and \ref{fig:ksps_SKM_MS42}d. Between spectra 50 and 100, the comparison mask is only 3-5 channels wide, while the MS-42 mask is immediately 7 channels wide, contributing a significant amount of false positives. In addition, the \SKs mask is 9 channels wide starting at around spectrum 140 while the comparison mask is only 7 channels wide at the most. The larger the multi-scale shape is, the more RFI is flagged, at the expense of overflagging on clean data.

\subsection{Signal Data Rate to Duty Cycle}
\label{res:ksps-dc}

Due to the \SKs algorithm's known analytical weakness to 50\% duty cycles, it is pertinent to test this weakness out in practice and find configurations of RFI mitigation schema that mitigate this weakness. These simulations were run with $M=512$, carrier frequency centered on a PFB channel, and with a duty cycle period of 1ms, which is slightly shorter than the time length of a single \SKs bin. As described in Section \ref{sec:data}, this can cause the effective duty cycle as the \SKs algorithm sees it to be slightly different from the given intrinsic duty cycle.

\begin{figure}[ht!]
    \centering
    \includegraphics[width=0.7\linewidth]{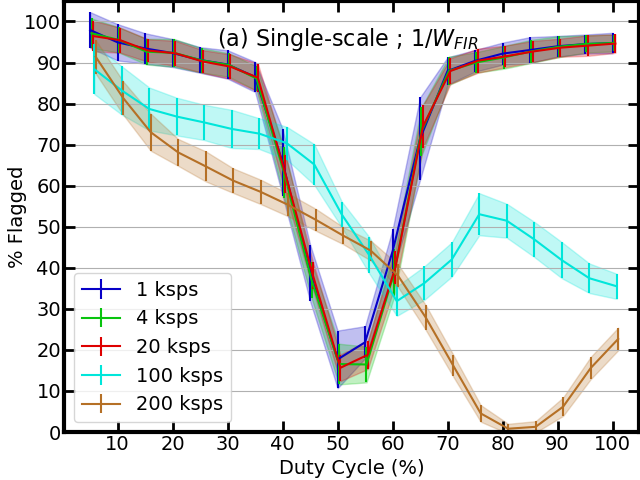}
    \includegraphics[width=0.7\linewidth]{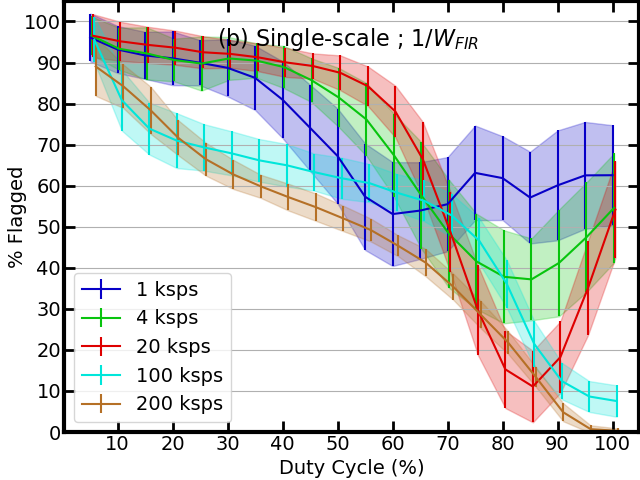}
    \caption{Flagging percentages for various data rates compared to the duty cycle of the RFI signal. BPSK on the top and 2-bit ASK on the bottom.\newline}
    \label{fig:ksps_dc_SS}
\end{figure}

As shown in Figure \ref{fig:ksps_dc_SS}a, low data rate BPSK signals are indeed minimally flagged at 50\% duty cycle. These signals are flagged at or around a 90\% rate except in the range of ~35\% to 70\% duty cycle. There are, however, some very clear flagging differences with higher data rates. In our simulated spectrometer setup, the 100 and 200 ksps signals are not sampled entirely accurately. For these data rates, the symbols change at a comparable rate to the spectra dump rate, which leads to unexpected \SKs flagging results. In Figure \ref{fig:ksps_dc_SS}a, we see that the minimum flagging occurs at 60\% and 80\% duty cycle for 100 and 200 ksps signals, respectively. For the 100 ksps signal, the center channel is not flagged at all at 60\% duty cycle, but the sidelobes get progressively less flagged as the duty cycle rises, leading to the local maximum at 80\%. These effects can be entirely attributed to the fast data rate since they can be eliminated by changing the spectrometer characteristics. When we multiply the sampling rate by a factor of 10 and reduce the duty cycle period by the same factor, the two faster data rate results look identical to the lower data rate results. A factor of 10 is chosen because that is the difference between 20 ksps and 200 ksps, the highest cleanly flagged data rate and the maximum data rate picked. It is also worth noting that the total power rises with duty cycle, as the signal spends more time turned on.

When using a less strict 4/\wfir\, frequency cutoff, the 20 ksps signal has some sidelobe emission not captured by single-scale \SK. The flagging rate is then slightly less than the 1 ksps and 4 ksps signals, and the difference becomes larger for higher duty cycles, up to 20\% less flagging. QPSK signals are flagged similarly to BPSK signals, although slightly less at higher data rates and higher duty cycles again due to the slightly larger sidelobes.

Amplitude-shift keyed signals have a bit of a different response, as shown in figure \ref{fig:ksps_dc_SS}b. For lower data rates, there is a weakness at around 85\% duty cycle instead of the 50\% of BPSK signals. This can be attributed to the fact that since the amplitude changes over time, the effective duty cycle of the signal becomes larger than the intrinsic duty cycle, where the intrinsic duty cycle is the percentage of time where the signal is physically on, and the effective duty cycle is when the signal appears to be on. This is explored more in Section \ref{sec:disc-dc}. With this result, we realize that \SKs is weak to 50\% effective duty cycle signals rather than 50\% intrinsic duty cycle.

For BFSK signals, there is also no minimally flagged duty cycle - signals with a data rate of 20 ksps or less start at $>$90\% for low duty cycle and flatten out at about 60\% flagged at higher duty cycles, while higher data rates continue to drop - for example, 200 ksps signals are 20\% flagged for 90\% duty cycle. The expected minimum flagging rate should happen at 100\% duty cycle, since the natural duty cycle of a single frequency channel is 50\%, owing to the fact that the data stream of a BFSK signal with random bits should be half 1's and half 0's. Thus, on a channel-by-channel basis, FSK signals act like on-off keyed (OOK) signals.

\begin{figure}[ht!]
    \centering
    \includegraphics[width=0.7\linewidth]{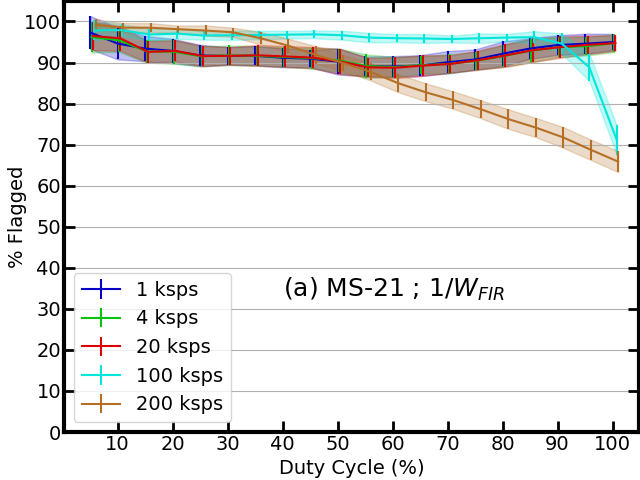}
    \includegraphics[width=0.7\linewidth]{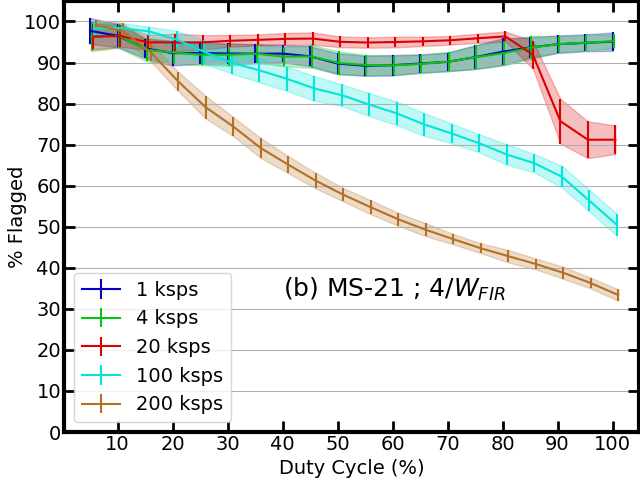}
    \caption{Flagging percentages for various data rates compared to the duty cycle of the RFI signal. Signals with a $1/W_{FIR}$ cutoff in the the symbol smoothing window are shown on the top and $4/W_{FIR}$ on the bottom. These signals are modulated with the BPSK scheme, $M=512$, and mitigated with MS-21 \SK.\newline}
    \label{fig:ksps_dc_chart-MS21}
\end{figure}

\begin{figure*}[ht!]
    \begin{center}
    \begin{tabular}{ccc}
        \includegraphics[width=0.32\linewidth]{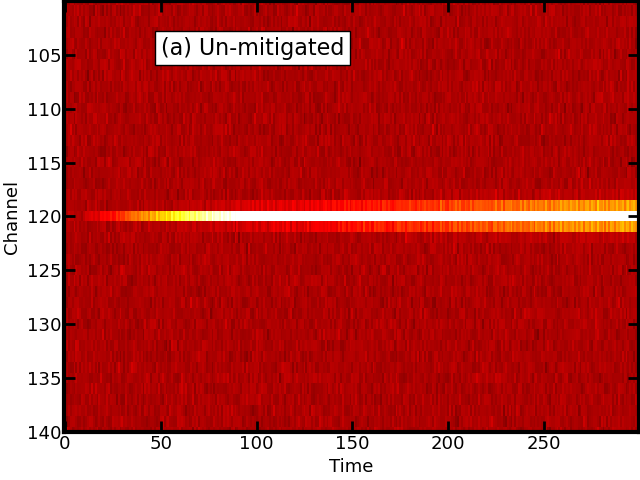}
    &
        \includegraphics[width=0.32\linewidth]{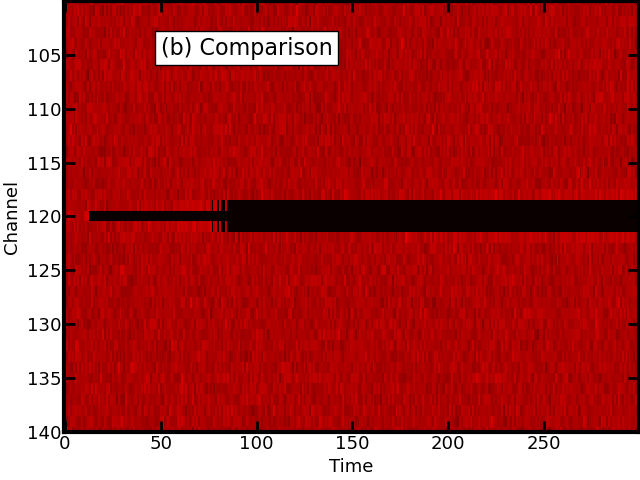}
    &
    \multirow{2}{*}[2in]{
    \includegraphics[width=0.06\linewidth]{misc/cbar.png} } \\
        \includegraphics[width=0.32\linewidth]{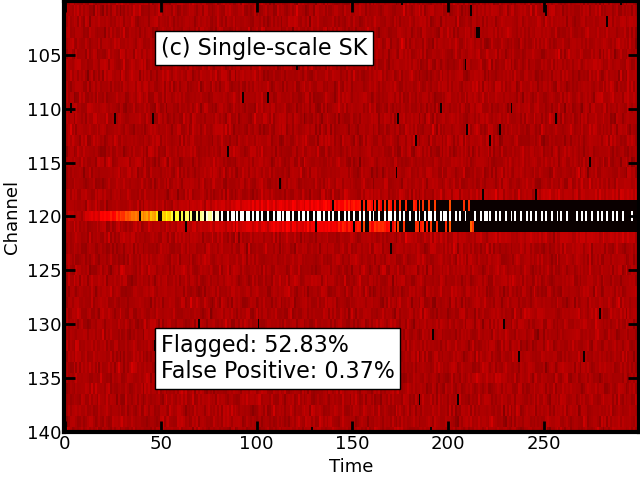}
    &
        \includegraphics[width=0.32\linewidth]{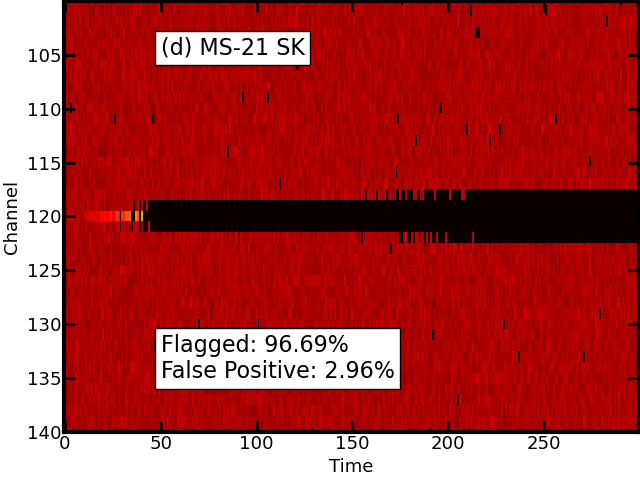}
    &
    \end{tabular}
    \end{center}
    \caption{Waterfall plots of unmitigated and mitigated RFI, for comparing MS-21 \SKs flagging to the duty cycle of the RFI signal. These signals are modulated with the BPSK scheme with 50\% duty cycle, centered at channel 120 and mitigated with $M=512$. (a) \& (b) Unflagged and flagged 1 ksps signal. (c) \& (d) 20 ksps signal. (e) \& (f) 100 ksps signal. (g) \& (h) 200 ksps signal.\newline}
    \label{fig:ksps-MS21-spectra-50}
\end{figure*}

In order to test if multi-scale \SKs can cover up some of the weaknesses of single-scale \SK, we run the same experiment as above with a MS-21 \SK. This bin shape was chosen as it is the smallest multi-scale \SKs shape that significantly improves flagging. Adding one extra channel in the multi-scale shape is indeed enough to detect the center channel when it gets missed by single-scale \SKs due to 50\% duty cycle, as shown in Figures \ref{fig:ksps_dc_chart-MS21} and \ref{fig:ksps-MS21-spectra-50}. The slow, constant drops in flagging for 100 ksps and 200 ksps signals can be attributed to stronger sidelobe emission as the duty cycle rises. The steep drops at $<$80\% duty cycle for the 100 ksps signal in Figure \ref{fig:ksps_dc_chart-MS21}a and the 20 ksps signal in Figure \ref{fig:ksps_dc_chart-MS21}b are due to the first sidelobe channel not being detected. We can also again see the lower flagging rates for high duty cycles demonstrated in general by the less strict sidelobe suppression in panel b. 

\subsection{Data Rate and Duty Cycle to multi-scale \SKs bin shape}
\label{sec:ksps-mssk}

\begin{figure*}[ht!]
    \centering
    \begin{tabular}{ c c }
    \includegraphics[width=0.32\linewidth]{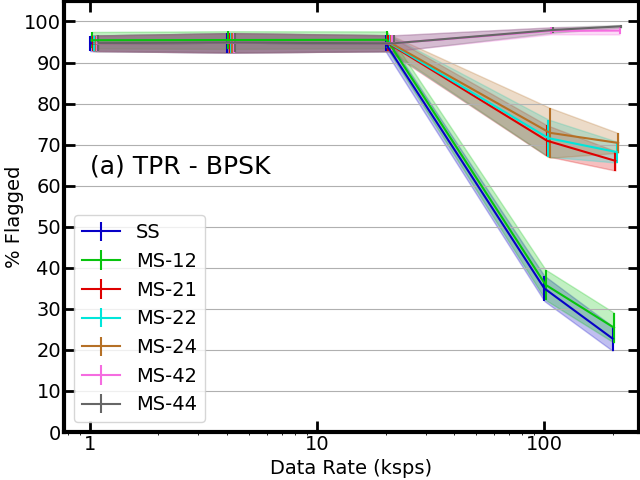}
    &
    \includegraphics[width=0.32\linewidth]{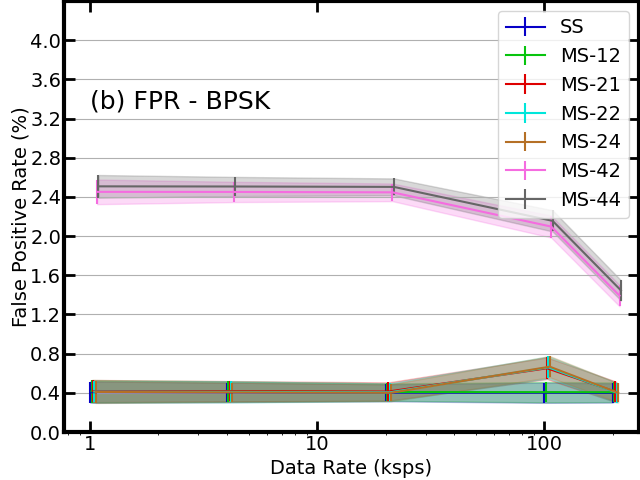}
    \\
    \includegraphics[width=0.32\linewidth]{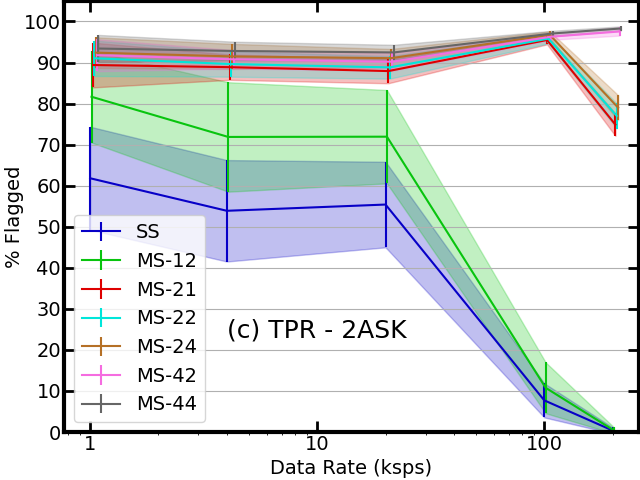}
    &
    \includegraphics[width=0.32\linewidth]{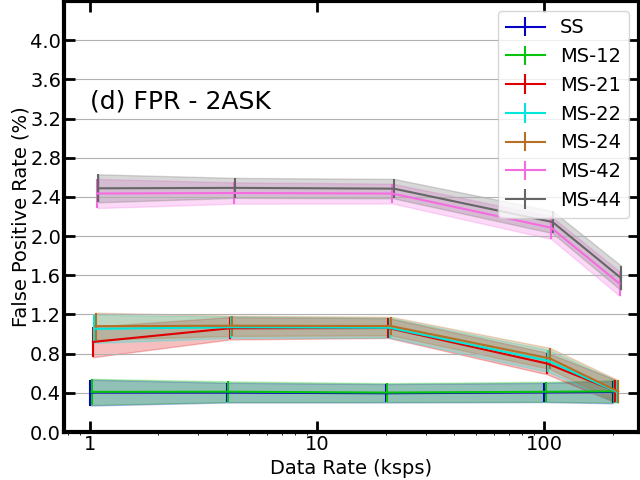}
    \\
    \end{tabular}
    \caption{Flagging percentages (top) and overflagging percentages (bottom) for single-scale and several multi-scale \SKs shapes compared to the data rate of the RFI signal. These signals are modulated with the BPSK scheme with 100\% duty cycle, centered in channel 120, and $M=512$.\newline}
    \label{fig:ksps_mssk__chart}
\end{figure*}

We also tested how different multi-scale \SKs bin shapes flagged at different data rates. Signals were generated with a 100\% duty cycle, carrier frequency centered on a PFB channel, and with $M=512$. From Figure \ref{fig:ksps_mssk__chart}a and b, we see that adding an extra channel to the MS-\SKs bin shape for MS-2X mitigation helps flag more data, and adding 3 more channels for MS-4X \SKs mitigation raises the flagging level above 90\% for all data rates. However, in the case of MS-4X, the FPR rises significantly above the theoretical lowest rates from Figure \ref{fig:fpr_noise}, up to 2.4\%. At these lower data rates, the multi-scale bin shape flags more channels than the width of the signal, leading to the high FPR rates. As the the signal data rate rises, the sidelobes become wider and more closely match the multi-scale \SKs width, explaining the drop in FPR at those points. Another thing to note for BPSK signals is that adding any extra time bin width, or subsequent \SKs bins in the same channel, does nothing to increase flagging. Single-scale and MS-12 \SKs perform identically, as do MS-21, MS-22, and MS-24 to each other, as well as MS-42 and MS-44.

Figure \ref{fig:ksps_mssk__chart}c and d show that adding an extra time bin in multi-scale \SKs does have an advantage for amplitude-keyed signals such as 2-bit ASK, but only if the multi-scale channel width is 1. MS-12 \SKs mitigation performs better than single-scale \SKs at lower data rates, but they both suffer the same lack of flagging at 200 ksps that was described in Section \ref{sec:ksps-m}. Adding channels increases flagging the same way as Figure \ref{fig:ksps_mssk__chart}a, but eliminates the effectiveness of adding extra time bins as well as increasing false positive rates. We can still see, however, that MS-2X still flags more RFI than MS-12.

We also test multi-scale \SKs shape against duty cycle. We find again that across the modulation types tested, the multi-scale channel width makes much more difference than the \SKs spectra bin width. MS-4X generally flags much better than MS-2X or single-scale \SK, but can contribute significant overflagging. The only relation to duty cycle found is that there is much less overflagging at higher duty cycles.


\subsection{Offset carrier frequencies}

\begin{figure*}[ht!]
    \centering
    \begin{tabular}{ c c c }
    \includegraphics[width=0.28\linewidth]{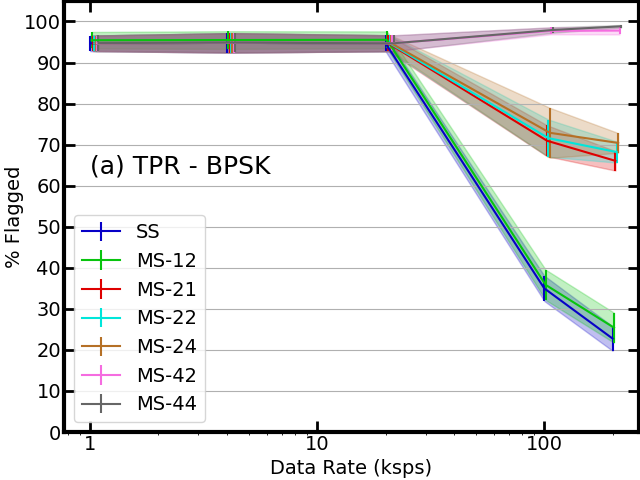}
    &
    \includegraphics[width=0.28\linewidth]{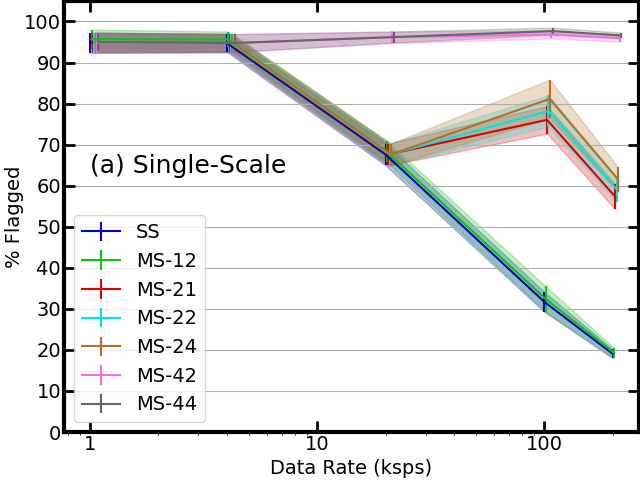}
    &
    \includegraphics[width=0.28\linewidth]{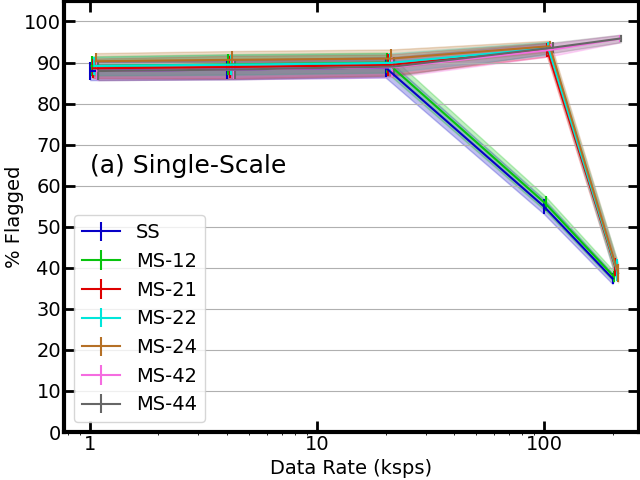}
    
    \\
    \end{tabular}
    \caption{BPSK signal flagging while the carrier frequency is (a) centered on a channel, (b) offset by a quarter channel, and (c) offset by a half channel. In all cases, MS-4X flags at above 90\%}
    \label{fig:offset}
\end{figure*}

We explored the case of when the signal carrier frequency doesn't line up with the center of a spectrometer channel and found some interesting differences in flagging. Every experiment above was redone with the signal offset by a quarter channel and a half channel, as there is no reason to expect a RFI transmitter carrier frequency to line up with a radio astronomical spectrometer channel. Results for a simple 20 ksps BPSK signal across duty cycles are shown in Figure \ref{fig:offset}. The differences from channel centered single-scale \SKs flagging can be explained by the central channel emission and sidelobe emission mixing, where we now know the former gets flagged better than the latter in most cases. However, we can see that multi-scale \SKs once again removes any weaknesses to the sidelobe emission.

\subsection{Astronomical Signals}

\begin{figure*}[ht!]
    \centering
    \begin{tabular}{ c c c }
    \includegraphics[width=0.32\linewidth]{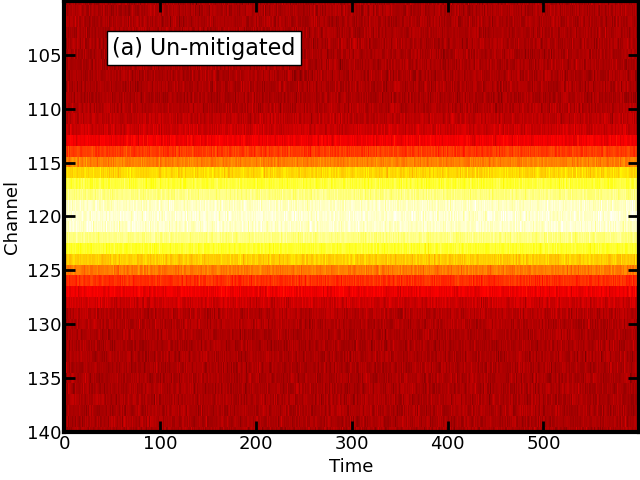}
    &
    \includegraphics[width=0.32\linewidth]{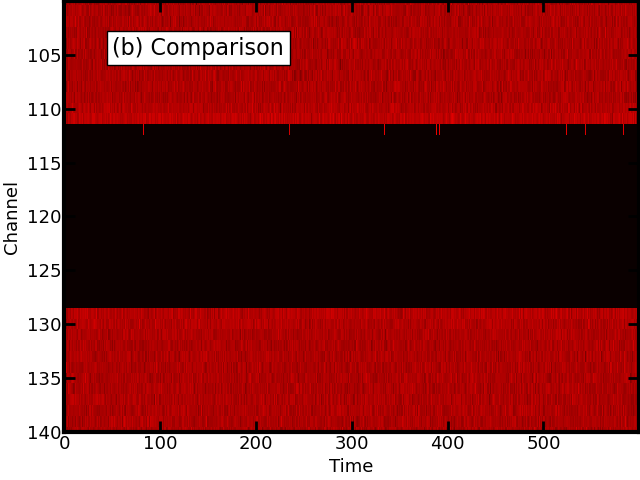}
    &
    \multirow{2}{*}[2in]{
    \includegraphics[width=0.06\linewidth]{misc/cbar.png} } \\
    \includegraphics[width=0.32\linewidth]{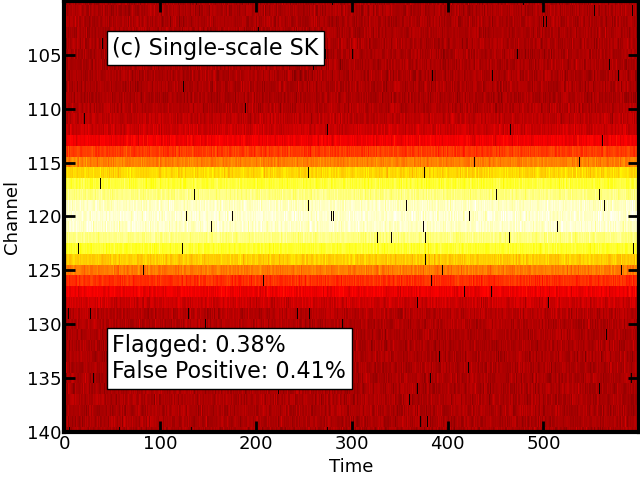}
    &
    \includegraphics[width=0.32\linewidth]{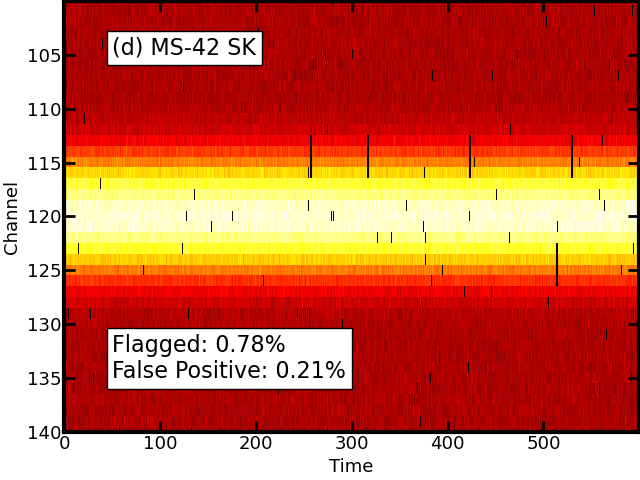}
    &
    \end{tabular}
    \caption{An incoherent stationary astronomical signal that escapes detection by both single-scale and MS-42 \SKs.}
    \label{fig:astrosig}
\end{figure*}

As \SKs has been advertised as allowing stationary Gaussian signals to pass through, we touch on injecting a basic astronomical signal and testing said claim. The results are shown in Figure \ref{fig:astrosig}. Comparing to Figure \ref{fig:fpr_noise}, we see the single-scale \SKs flags less than 0.4\%, which is the baseline false positive rate for $M=512$. MS-42 \SKs only flags 0.78\%, and seems to detect a few pixels along the edge of the signal. It is also worth noting that the signal here is much stronger than typical intensities at this un-averaged scale. Most observations of astronomical signals detections have to be averaged over 5 minutes or more to have a significant signal-to-noise. This is definitive evidence that the \SKs estimator can detect and remove RFI while leaving scientific signals untouched, which is a large benefit for observations where the scientific signals of interest are mixed between both narrowband and wideband RFI. However, overlaying RFI on top of the scientific signal is beyond the scope of this paper, as it introduces a significant amount of extra dimensions depending on the RFI shape, the relative signal strengths, and the amount of overlap. This experiment is also only under the scope of incoherent astronomical sources. Coherent sources, such as masers, may be flagged by \SK, which we leave for future work.

\section{Discussion}
\label{sec:disc}

We discuss the results of the previous section and provide some deeper insight as to why certain signal characteristics were flagged at different levels.


\subsection{Duty Cycle}
\label{sec:disc-dc}

To expand on some of the differences in flagging across duty cycle and for different modulation types, we restate equation \ref{eq:var_mean2}, from which the \SKs estimator is derived:

\begin{equation}
V_k = \frac{\sigma_k^2}{\mu_k^2}
\end{equation}

\begin{figure*}[ht]
    \centering
    \begin{tabular}{ c c }
        \includegraphics[width=0.4\linewidth]{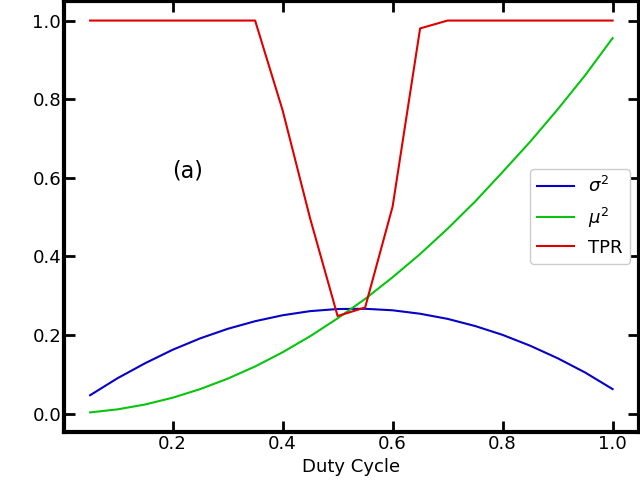}
        &
        \includegraphics[width=0.4\linewidth]{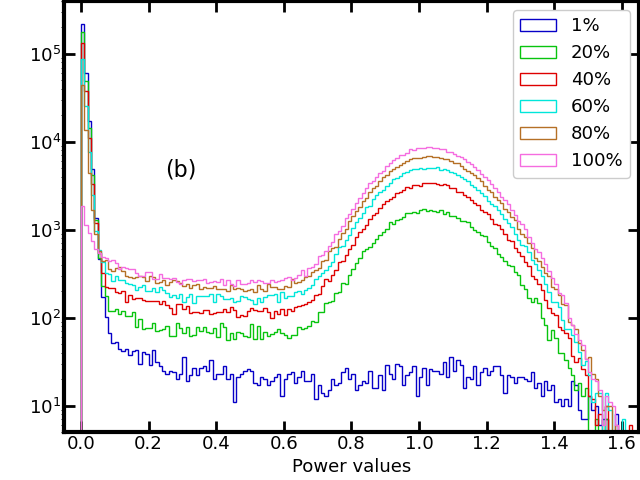}
        \\
        \includegraphics[width=0.4\linewidth]{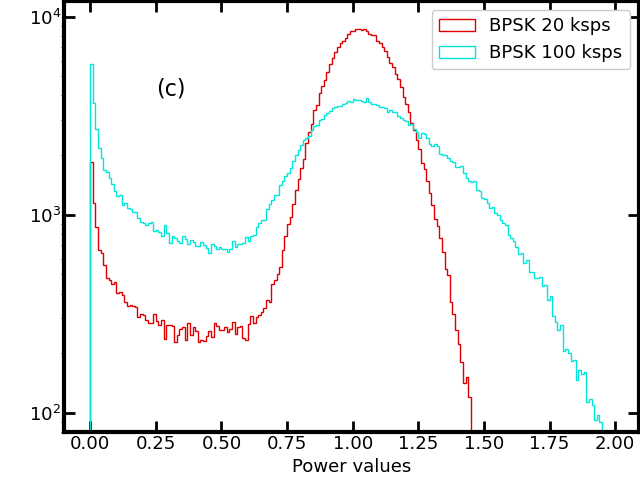}
        &
        \includegraphics[width=0.4\linewidth]{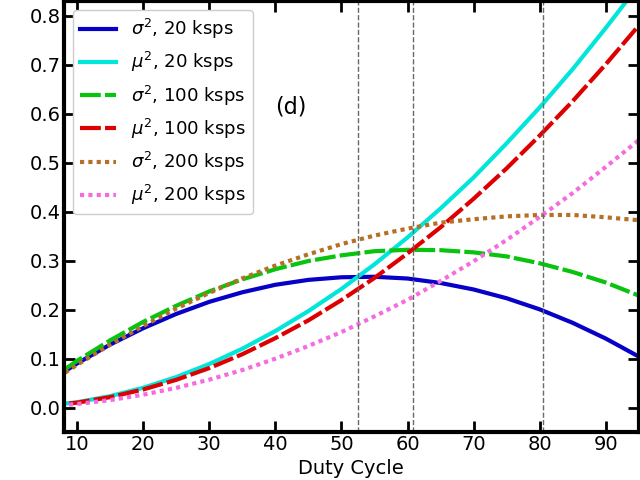}
    \end{tabular}
    \caption{A glance at some statistical properties of a BPSK signal. (a) Comparing $\mu^2$ and $\sigma^2$ in the channel center of the RFI with the TPR rate. (b) Power distributions in the central channel for a range of duty cycles. (c) Power distributions in the central channel for two different data rates. The 100 ksps signal is undersampled. (d) Comparing $\mu^2$ and $\sigma^2$ in the central channel for three different data rates. The vertical dashed lines mark where $\mu^2$ and $\sigma^2$ cross each other for each data rate.}
    \label{fig:bpsk_mv1}
\end{figure*}

We will look into comparing the squared means and variances as well as histograms of power values to give a deeper look into the disparities in TPR for various RFI characteristics. For the simplest case, we show the squared mean and variance for the center channel of a 20 ksps BPSK signal as well as the flagging rate in Figure \ref{fig:bpsk_mv1}a. As the mean and variance approach each other, $V_k$ becomes 1 and the signal does not get flagged. As for the shapes of the lines, we refer to Figure \ref{fig:bpsk_mv1}b, which shows the histograms of power values at several select duty cycles. These signals are generated with 600 \SKs bins instead of 60 to decrease statistical error and with no ramp-up in strength; they are at full power for the length of the simulation. 

As the duty cycle rises, the signal spends more time turned on and the overall power in the center channel rises. To explain the shape of the variance trend, notice that at low duty cycles, most of the values reside at lower power - so the variance is low. As the duty cycle rises, more of those values are shifted towards a distinct population of higher power values, and the variance rises. At some point, there are more values at higher power, and the variance starts to drop again. The relation of $\sigma^2$ to $\mu^2$ gives insight to when the signal is detected or not.

Looking back at Figure \ref{fig:ksps_dc_SS}a, the null in flagging for higher data rates occurs at a higher duty cycle, and Figure \ref{fig:bpsk_mv1}c and \ref{fig:bpsk_mv1}d tell us why. At 100ksps, the BPSK signal starts to have constructive and destructive interference inside the PFB, as the signal itself changes phase by 180 degrees on the same timescales as it gets added together. Constructive interference raises the effective signal power while destructive interference lowers it, resulting in spread out power values. This is reflected in Figure \ref{fig:bpsk_mv1}c, where the higher power population is spread out. As this happens, the mean of all power values gets lower, and the variance rises. This results in $\mu^2$ and $\sigma^2$ meeting each other at a higher duty cycle and pushing the null in flagging higher, which is shown in panel d). This only gets more dramatic for 200 ksps, where the null in flagging happens at even higher duty cycles.

The \SKs detection of ASK signals can also be analyzed in the same way. Looking at the TPR of signals in Figure \ref{fig:ksps_dc_SS}b, the local minima do not occur at duty cycles of 50\%. The reason is that the effective duty cycle (what the spectrometer sees per channel) for BFSK and BASK signals is different than the intrinsic duty cycle (when the transmitter is on or off). On a channel-by-channel basis, a BFSK signal acts like and on-off keyed (OOK) signal, so in a single channel with a 100\% intrinsic duty cycle and across a large amount of random bits, the channel occupancy should be 50\%, leading to an effective duty cycle of 50\%. Thus, single-scale \SKs is weak to BFSK signals with intrinsic duty cycles of 100\% unless the bit rate is slow enough that the binary distribution of 1's and 0's cannot be expected to be half and half in a single \SKs time bin of power values.

The effective duty cycle of ASK signals is likewise affected by the power modulation. Since a portion of the signal is by definition at a lower amplitude, the drop in power can appear as if the signal is not on for as long.

As we can see in Figure \ref{fig:bpsk_mv1}a, the squared mean and variance intersect for the BPSK signal at duty cycles of 0\% (no RFI) and 50\%. When the two quantities are equal to each other, $V_k = 1$ and \SKs does not detect RFI. For the BPSK signal, the effective duty cycle matches the intrinsic duty cycle and we get the characteristic weakness at 50\%. However, for the BASK signal, we can see that the effective duty cycle is about half of the intrinsic duty cycle - the squared mean and variance approach each other at half the rate of the BPSK signal, only matching up when the intrinsic duty cycle is 100\%. BFSK signals have the same effect.

\begin{figure*}[ht!]
    \centering
\begin{tabular}{ c c }
    \includegraphics[width=0.4\linewidth]{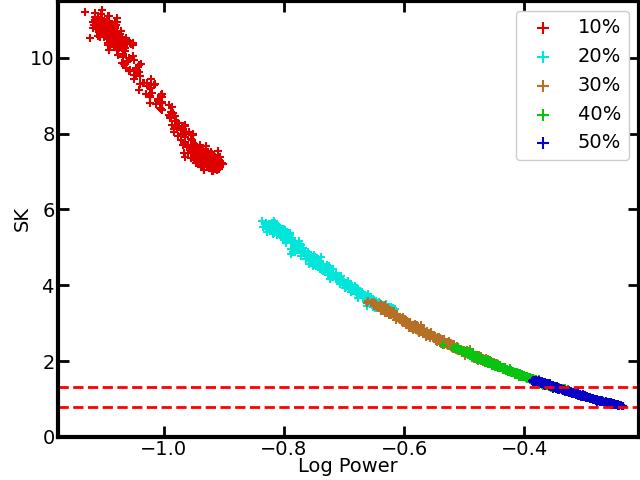}
    &
    \includegraphics[width=0.4\linewidth]{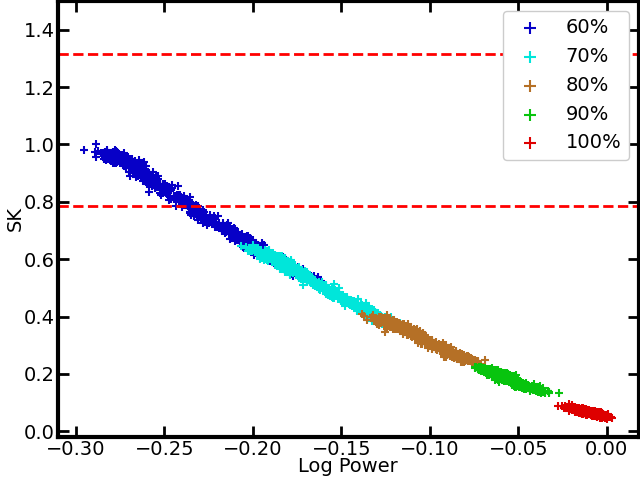}
\end{tabular}
\caption{Scatter plots of BPSK signals at various duty cycles comparing average power to \SKs value in the center channel. The data is split into 2 separate graphs for readability: 10\% to 50\% duty cycle on the left, and 60\% to 100\% on the right. Note how the data lies above 1 for duty cycles less than 50\%, approaches \SKs values of 1 as the duty cycle approaches 50\%, and goes below 1 for higher duty cycles. \SKs thresholds are outlined in red.}
\label{fig:bpsk-ss-dc-scatter}
\end{figure*}
\begin{figure*}[ht!]
    \centering
\begin{tabular}{ c c }
    \includegraphics[width=0.4\linewidth]{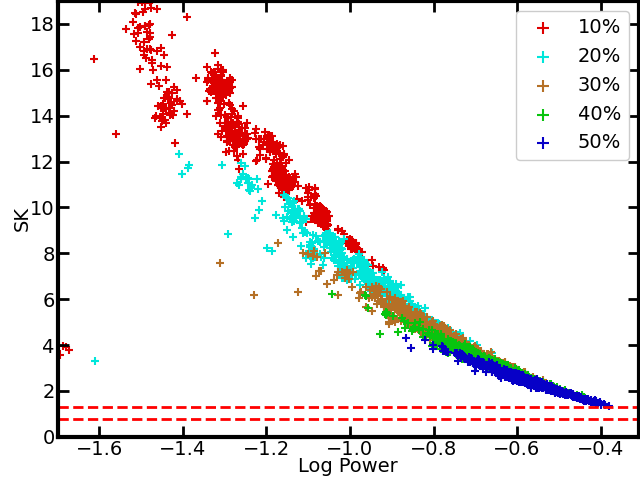}
    &
    \includegraphics[width=0.4\linewidth]{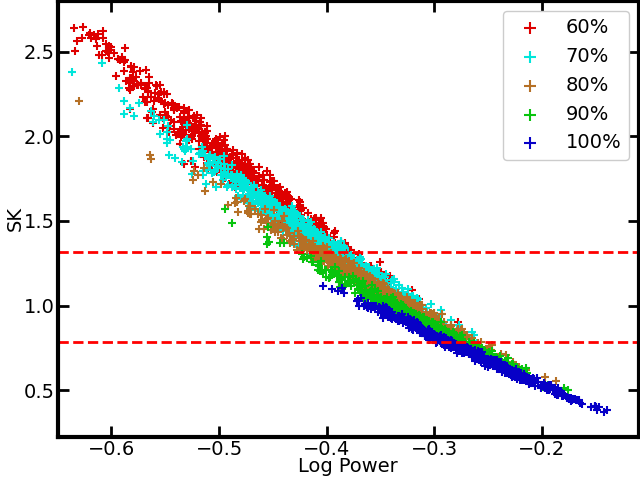}
\end{tabular}
\caption{Scatter plots of BASK signals at various duty cycles comparing average power to \SKs value in the center channel. The data is split into 2 separate graphs for readability: 10\% to 50\% duty cycle on the left, and 60\% to 100\% on the right. For these signals, a 100\% intrinsic duty cycle acts as a 50\% effective duty cycle, and doesn't get flagged. \SKs thresholds are outlined in red.}
\label{fig:ask-ss-dc-scatter}
\end{figure*}

Figures \ref{fig:bpsk-ss-dc-scatter} and \ref{fig:ask-ss-dc-scatter} compare the log average power vs. \SKs of BPSK and BASK signals at a wide range of duty cycles, from 10\% to 100\%. We see for the BPSK signal how signals at 10\% and 100\% are flagged as they lie outside of the acceptable \SKs ranges denoted by the red dashed lines. However, the 50\% duty cycle signal lies almost entirely within the acceptable ranges and does not get flagged. On the other hand, the BASK signals in Figure \ref{fig:ask-ss-dc-scatter} at 100\% are not flagged as their effective duty cycle is 50\%. The distinct lines seen at low duty cycles are an effect of the small amount of RFI present in the signal - only a few bits of data get through in each \SKs bin, leading to a low number of discrete relative power levels as the signal ramps up in intensity.

\subsection{Multi-scale \SK}
\label{sec:disc-mssk}

It was found that applying multi-scale \SKs on top of the existing single-scale \SKs helps cover some of the weaknesses to duty cycle and data rate. The most striking conclusion is that including more than one channel ($m>1$) in the multi-scale \SKs bin shape is much more beneficial than including more than one concurrent single-scale \SKs bin ($n>1$) in the same channel, for the RFI signals tested. Generally, multi-scale \SKs bin shapes with $n>m$ flag just as well as $n\le m$. So, it is most advantageous to have at least $m \ge 2$ for effective multi-scale \SKs. However, this can change significantly based on the RFI environment and the frequency resolution of the spectrometer. For higher data rates or higher frequency resolution, $m>2$ is suggested, but significant overflagging can occur if $m$ is greater than the typical channel width of RFI. Depending on the expected RFI environment, the frequency width of the spectrometer, and how flagged data is treated, the user may opt for higher overflagging if it means a more robust flagging mask. This is not dissimilar to many other RFI mitigation techniques, where a more rigorous mask can lead to clean data being flagged. This may also be a symptom of the \SKs mask picking up on RFI missed by the comparison mask, which is a good thing.

All of the tests described in section \ref{sec:results} involve a duty cycle period far shorter than the amount of time it takes to collect $M$ spectra which means that, for a signal with a duty cycle smaller than 100\%, a single \SKs bin sees the signal turn on and off many times. Multi-scale \SKs with higher $n$ was also tested in the case of discontinuous RFI - where the signal is 100\% on in one \SKs bin and off in the next. In this case, a multi-scale shape of $m=1,n=2$ will see two adjacent data bins with duty cycles of 100\% and 0\%, the duty cycle of which is 50\%, following equation \ref{eq:ms_dc}. Multi-scale \SKs would not contribute any flagging on the central channel in this case, though it may continue to find more side-band interference.

\subsection{Data Rate}

In general, the higher the data rate went, the harder it was for \SKs to flag it. This is true for both single scale and any of the multi-scale \SKs shapes tested. Given a constant signal power level, as the data rate goes up, more power leaks into the sideband channels relative to the central channel. As described in Section \ref{sec:ksps-m}, the modulation of the carrier wave acts like a square wave in time. As the data rate rises, the period of said square wave lowers, which in turn widens the resultant signal after being channelized via poly-phase filterbank. Sidelobe emission starts to become significant and does not get flagged optimally, leading to the result that the larger the data rate is, the less likely the full signal will be flagged.

This result should be taken into context with frequency resolution. If we use a coarser spectrometer resolution with the same high data rate signal such that the sidelobe emission can be contained to one channel, flagging will improve. This was proven in our simulations by increasing the sample rate and stretching the same amount of channels over a wider bandwidth.
 
It is unclear how the statistical properties of sidelobe emission cause it to be missed by single-scale \SK. It is clear from Figures \ref{fig:ksps_SKM_MS42}c that some sidelobe emission gets flagged, but most does not. We followed the same methods as described in Section \ref{sec:disc-dc} to track $\sigma^2$ and $\mu^2$, which resulted in no clear explanation.

The statistical distributions of power values in sidelobe channels were found to have different shapes from those in the center channel, so while the sidelobes are certainly at lower power, there is something else causing the non-detection. This can be proven by Figure \ref{fig:ksps_SKM_MS42}c, where we see bright non-detected sidelobes and dimmer detected center channel emission. However, we do see that using multi-scale \SKs with enough frequency width to cover the sidelobe emission can detect the entire signal, comparing panels c and d of Figure \ref{fig:ksps_SKM_MS42}. Despite not knowing why single-scale \SKs does not detect this emission, we can see that multi-scale \SKs does nullify this weakness even in the case that the center channel doesn't get flagged, as we elaborated on in Section \ref{sec:disc-mssk}.

\subsection{Signal strength}

The RFI signals were ramped up in strength in order to ascertain what signal-to-noise ratio (S/N) was required for \SKs to capture the signal, if any. To do this, we record the first instance in time of the signal being flagged and the last instance of the signal being unflagged. A signal-to-noise ratio was computed by dividing the arbitrary signal strength by one standard deviation of the noise at the two time indices. Using Figure \ref{fig:ksps_SKM_1}a as an example, the central channel becomes completely flagged between \SKs bins 20 and 26, where the S/N ranges from 14.9 to 23.3. This process would run 100 times and averages of the bin indices and S/N ratios was recorded. 

This was done for BPSK signals at various data rates and with different noise levels. For each data rate, the noise level was gradually raised. It was found that the central channel goes from first being flagged to being completely flagged at the same S/N ratios, regardless of noise power level. A 1 ksps BPSK signal with 100\% duty cycle starts getting flagged at a S/N of ~15 and gets completely flagged at a S/N of ~24. A 200ksps signal starts getting flagged at a S/N of ~19 and gets completely flagged at a S/N of ~44. It seems, therefore, that higher data rate signals are flagged at a higher minimum S/N ratio than lower data rate signals when the duty cycle is 100\%. It is not clear why lower S/N signals are not flagged. This can perhaps be mitigated by running MS-\SKs with a large amount of time bins or using a large value of $M$, but these can be subjective based on data set sizes and RFI environment. While we recommend best practices for \SKs usage in this paper, it is always worth testing some of the mitigation parameters in a real data set to make sure RFI is getting adequately flagged.



\subsection{Digitization and Truncation}

We recreated the negative bias in \SKs values due to truncated digitization found by \citet{mirhosseini2020high}, where the signal saturates the analog-digital converter. The more saturated the digitizer becomes, the lower \SKs values are returned. This does have an effect on single-scale \SKs tests, especially when comparing against duty cycle. As shown in Figure \ref{fig:bpsk-ss-dc-scatter}, expected \SKs values run from above 1 to below as the duty cycle increases, so a negative bias shifts the null in flagging to lower duty cycles. However, multi-scale \SKs is not affected by this bias and still flags at the same higher rates regardless of the amount of truncation.

\section{Conclusions}

We implemented single-scale and multi-scale \SKs in Python and applied this RFI mitigation method to simulated data with realistic RFI signals. The signals were modulated using several basic encoding schemes, with various parameters that included data rate, duty cycle, and carrier frequency. We applied \SKs with different accumulation lengths and multi-scale bin shapes to determine how the RFI mitigation method responds to various signal types. We generated a comparison mask by converting the signal data to decibels using the noise as reference and setting RFI to be wherever the data is $<$ -10 dB. Using this mask, we then derived true and false positive rates. Results across different \SKs parameters and signal characteristics are summarized below.

\begin{itemize}
    \item Signals with higher data rates are typically harder to flag. In most cases, signals with data rates $>$100 ksps were flagged non-optimally, due to single-scale \SKs missing sidelobe emission.
    \item Signals that were encoded without any power change in a single channel (BPSK \& QPSK) were poorly flagged when they had an intrinsic 50\% duty cycle. However, signals with changes in power level (BFSK \& BASK) were poorly flagged at an intrinsic duty cycle greater than 50\%, explained by power modulation making the signals appear with an effective duty cycle that appeared to be 50\%, despite the intrinsic duty cycle being higher.
    \item We implemented multi-scale \SKs in addition to single-scale \SK, with various shapes $m\times n$ in the frequency and time dimensions. We found that $n>1$ only contributed to flagging in the case that $m=1$ and the signal had some form of amplitude modulation. Otherwise, $n$ did not contribute significantly to either flagging or overflagging.
    \item On the other hand, $m\geq2$ did help to allay some of the weaknesses to high data rate and duty cycle, which included flagging sideband spillover more effectively. $m=4$ contributed a significant amount of false positives as the signal's width did not exceed 4 channels in many cases, leading to an \SKs mask that was wider than the comparison mask.
    \item A stationary incoherent astronomical signal passes through both single-scale and multi-scale \SKs without being flagged, which shows the \SKs estimator is capable is discerning incoherent real signals of interest from RFI.
\end{itemize}

For future work, more modulation types can be implemented and tested, as we only chose a few of the simpler modulation types, and the list of possible modern encoding schemes contains lots of possible alternatives. These include minimum-shift keying, spread-spectrum modulation, and multiplexing types such as frequency division multiplexing, among others. The modulation types we explored in this paper can also be scaled up to larger amounts of bits per symbol such as 16-PSK, which can transmit 16 bits for every symbol instead of just 1. Improvements on spectral kurtosis for RFI mitigation can also be explored using \citet{barszcz2011novel,vass2008avoidance,wang2013energy}, as well as combining \SKs with other statistical methods such as Inter-Quartile Range Mitigation \citep{morello2022iqrm}.

\SKs is also a simple enough calculation to include online inside a spectrometer, and Smith et al. (2022, in preparation) test various \SKs parameters against pseudo-real time data taken with the GBT. The data is not averaged down and so we are able to try out different RFI mitigation configurations on the same data set and compare results. Our python code for running Spectral Kurtosis on custom FITS-like files can be found on Github.\citep{etsmit2022rfi}\footnote{\url{https://github.com/etsmit/RFI_Simulations}}. These files consist of a series of blocks that contain an informational header followed by 8-bit complex voltages that can be arranged into a 3-dimensional spectrogram array with dimensions of frequency, time, and polarization.

\section{Acknowledgements}

This work is supported by the National Science Foundation through
Advanced Technologies and Instrumentation grant \#1910302.  This
material is based upon work supported by the Green Bank Observatory
which is a major facility funded by the National Science Foundation
operated by Associated Universities, Inc.  ETS and DJP thank the West Virginia University Eberly College Dean's Office for partial support of this project.  DJP is supported through the South African Research Chairs Initiative of the Department of Science and Technology and National Research Foundation.  We also acknowledge advice from conversations with David McMahon and Jason Ray, as well as invaluable guidance on realistic signal generation from Dr. Chad Spooner.

\bibliography{references}
\bibliographystyle{aasjournal}

\end{document}